\documentclass[12pt]{article}  

\usepackage{epsfig}
\usepackage{graphicx}
\usepackage{amsmath}
\usepackage{amssymb}
\usepackage{ulem}
\usepackage{mdwlist}
\usepackage{color}
\usepackage{caption}

\usepackage{a41}
\usepackage{color}
\usepackage{cite}
\usepackage[rflt]{floatflt}
\usepackage{float}
\usepackage{slashed}

\newcommand{\MS}{\overline{{\sf MS}}}

\usepackage{array}

\usepackage[english]{babel}
\usepackage[latin1]{inputenc}
\usepackage[T1]{fontenc}
\usepackage{ae}

\usepackage{url}

\usepackage{amsmath, amsthm, amssymb}
\newtheorem{thm}{Theorem}[section]

\newtheorem{definition}[thm]{Definition}

\newcommand{\NN}{\nonumber}

\newcommand{\Li}{{\rm Li}}

\newcommand{\ep}{\varepsilon}

\usepackage{rotating}

\usepackage{graphicx}

\newcounter{mmacnt}
\def\restartmma{\setcounter{mmacnt}{0}}
\restartmma \catcode`|=\active
\def|#1|{\mathrm{#1}}
\catcode`|=12
\newenvironment{mma}{
 \par\smallskip
 \catcode`|=\active
 \parskip=0pt\parindent=0pt 
 \small
 \def\In##1\\{%
   \def\linebreak{\hfill\break\null\qquad}%
   \refstepcounter{mmacnt}
   \hangindent=2.5em\hangafter=0
   \leavevmode
   \llap{\tiny\sffamily In[\arabic{mmacnt}]:=\kern.5em}%
   \mathversion{bold}\footnotesize$\displaystyle##1$\normalsize
   \mathversion{normal}\par
 }%
 \def\Print##1\\{%
   \def\linebreak{\hfill\break}%
   \hangindent=2.5em\hangafter=0
   \leavevmode ##1\par}%
 \def\Out##1\\{%
   \def\linebreak{$\hfill\break\null\hfill$}%
   \kern\abovedisplayskip\par
   \hangindent=2.5em\hangafter=0
   \leavevmode
   \llap{\tiny\sffamily Out[\arabic{mmacnt}]=\kern.5em}
   \footnotesize$\displaystyle##1$\normalsize\hfill\null\par
   \kern\belowdisplayskip
 }%
 \def\Warning##1##2\\{%
   \def\linebreak{\hfill\break}%
   \hangindent=2.5em\hangafter=0
   \leavevmode
   {\scriptsize##1 : ##2}\par}%
}{%
 \par\smallskip
}

\usepackage{color}

\newenvironment{fshaded}{%
\MakeFramed {\FrameRestore}
}%
{\endMakeFramed}

\allowdisplaybreaks[4]

\begin{document}
\setlength{\baselineskip}{0.515cm}
\sloppy
\thispagestyle{empty}
\begin{flushleft}
DESY 19--216
\hfill 
\\
DO--TH 19/02\\
TTP 19--043\\
MSUHEP-19-026\\
SAGEX 19-30\\
December 2019\\
\end{flushleft}

\mbox{}
\vspace*{\fill}
\begin{center}

{\LARGE\bf  The three--loop single mass polarized}

\vspace*{3mm}
{\LARGE\bf pure singlet operator matrix element}

\vspace{4cm}
\large
J.~Ablinger$^a$, A.~Behring$^b$, J.~Bl\"umlein$^c$, A.~De Freitas$^c$, A.~von~Manteuffel$^d$,
\newline 
C.~Schneider$^a$ and K.~Sch\"onwald$^{b,c}$ 

\vspace{1.5cm}
\normalsize
{\it $^a$~Research Institute for Symbolic Computation (RISC),\\
                          Johannes Kepler University, Altenbergerstra\ss{}e 69,
                          A--4040, Linz, Austria}\\

\vspace*{3mm}
{\it $^b$~Institut f\"ur Theoretische Teilchenphysik},\\
{\it Karlsruher Institut f\"ur Technologie (KIT) D-76128 Karlsruhe, Germany}
\\

\vspace*{3mm}
{\it  $^c$ Deutsches Elektronen--Synchrotron, DESY,}\\
{\it  Platanenallee 6, D-15738 Zeuthen, Germany}
\\

\vspace*{3mm}
{\it $^d$~Department of Physics and Astronomy, \\ Michigan State University,
East Lansing, MI 48824, USA}
\\


\end{center}
\normalsize
\vspace{\fill}
\begin{abstract}
\noindent
We calculate the massive polarized three--loop pure singlet operator matrix element $A_{Qq}^{(3), \rm PS}$
in the single mass case in the Larin scheme. This operator matrix element contributes to the massive
polarized three--loop Wilson coefficient $H_{Qq}^{(3),\rm PS}$ in deep--inelastic scattering and constitutes 
a three--loop transition matrix element in the variable flavor number scheme. We provide analytic results
in Mellin $N$ and in $x$ space and study the behaviour of this operator matrix element in the region of 
small and large values of the Bjorken variable $x$.
\end{abstract}

\vspace*{\fill}
\noindent
\numberwithin{equation}{section}
\newpage

\section{Introduction}
\label{sec:1}

\vspace{1mm}
\noindent
Higher order heavy flavor corrections to deeply--inelastic structure functions are important both in the unpolarized 
and polarized case \cite{Dittmar:2005ed,Boer:2011fh}. Their scaling violations are different if compared to the massless 
case and, therefore, influence the measurement of the strong coupling constant $\alpha_s(M_Z)$ from the structure functions
\cite{Bethke:2011tr,Moch:2014tta,Alekhin:2016evh}. Related to it, the massless parton distributions are unfolded, requiring a 
correct description of the heavy flavor effects. On the other hand, in order to describe the transition of massive partons 
becoming 
effectively massless, 
the variable flavor number scheme can be used \cite{Buza:1996wv,Ablinger:2017err,Blumlein:2018jfm}. This transition is described by 
massive operator matrix elements (OMEs), and after its application, effective calculations for scattering reactions at hadron 
colliders are 
possible, based also on heavy quark parton distributions. Several of these transition matrix elements have been already computed
in the unpolarized and polarized case in the single,  
cf.~\cite{Bierenbaum:2009mv,Ablinger:2010ty,Ablinger:2014lka,Ablinger:2014nga,Behring:2014eya,Ablinger:2014vwa,
Ablinger:2014uka,Ablinger:2017ptf}, and two--mass case  
\cite{Ablinger:2011pb,Ablinger:2012qj,Ablinger:2017err,Blumlein:2018jfm,Ablinger:2018brx,Ablinger:2019gpu}.

In this paper we calculate the massive polarized three--loop pure singlet operator matrix element $A_{Qq}^{(3), \rm PS}$ in the 
single mass case. The corresponding two--mass corrections, which require different computational techniques, have been 
computed in 
Ref.~\cite{Ablinger:2019gpu}. In the present calculation similar techniques as in Ref.~\cite{Ablinger:2014nga} are used. 
This has become possible upon finding the correct projector in the case of external massless fermion lines in 
\cite{Behring:2019tus}, which differs from the one 
in \cite{Buza:1995ie}. The main quantity to be derived is the constant part of the 
unrenormalized polarized massive pure singlet OME, $a_{Qq}^{\rm PS, (3)}$.

The paper is organized as follows. In Section~\ref{sec:2} we outline the basic formalism and give an overview of the 
calculation. In Section~\ref{sec:3} we present the results for $a_{Qq}^{\rm PS, (3)}$ and the massive OME $A_{Qq}^{\rm PS, 
(3)}$. Both quantities are given in $N$ and $x$ space and we discuss numerical results for  $a_{Qq}^{\rm PS, (3)}(x)$.
Section~\ref{sec:4} contains the conclusions. In an ancillary file to this paper we give the massive OME in computer 
readable form.

\section{Basic Formalism and Overview of the Calculation}
\label{sec:2}

\vspace{1mm}
\noindent
The pure singlet massive operator matrix element describes the transition between massless on--shell quark states 
$\langle q|$ in association with a local quark operator in the light-cone expansion \cite{LCE}, which in general is 
either located on the heavy quark line or on a massless quark loop. The latter case is denoted by $A_{qq,Q}^{\rm PS}$ and 
contributes to heavy quark corrections in case of massless final states only, which will be presented elsewhere because 
of the different context. In this paper we present the results for $A_{Qq}^{\rm PS}$. The first contribution to 
$A_{Qq}^{\rm PS}$ arises at two loops. Therefore, the corresponding expansion in the strong coupling constant $\alpha_s$ 
is given by, cf.~\cite{Bierenbaum:2009mv},
\begin{eqnarray}
A_{Qq}^{\rm PS} &=& 
  a_s^2 A_{Qq}^{(2),\rm PS}
+ a_s^3 A_{Qq}^{(3),\rm PS}
+O(a_s^4)~,
\end{eqnarray}
where $a_s = g_s^2/(4\pi)^2 \equiv \alpha_s/(4\pi)$. 
We perform the calculations in $D=4+\ep$ dimensions, leading to the following
pole structure of the unrenormalized OME at two- and three--loop order

\begin{eqnarray}
\hat{A}_{Qq}^{(2), \rm PS} &=& \left(\frac{\hat{m}^2}{\mu^2}\right)^\ep 
\biggl(
-\frac{\hat{\gamma}_{qg}^{(0)} \gamma_{gq}^{(0)}}{2 \ep^2}
+\frac{\hat{\gamma}_{qq}^{(1), \rm PS}}{2 \ep}
+a_{Qq}^{(2), \rm PS}+\ep \bar{a}_{Qq}^{(2), \rm PS}
\biggr), \\
\hat{A}_{Qq}^{(3), \rm PS} &=& \left(\frac{\hat{m}^2}{\mu^2}\right)^{3 \ep/2} 
\Biggl\{
\frac{\hat{\gamma}_{qg}^{(0)} \gamma_{gq}^{(0)}}{6 \ep^3} \big(
\gamma_{gg}^{(0)} - \gamma_{qq}^{(0)} + 6 \beta_0 + 16 \beta_{0,Q}
\big)
\nonumber \\ && \phantom{\left(\frac{\hat{m}^2}{\mu^2}\right)^{3 \ep/2} \biggl\{ }
+\frac{1}{\ep^2} \biggl[
-\frac{4}{3} \hat{\gamma}_{qq}^{(1), \rm PS} \big(\beta_0+\beta_{0,Q}\big) - \frac{1}{3} \gamma_{gq}^{(0)} \hat{\gamma}_{qg}^{(1)}
\nonumber \\ && \phantom{\left(\frac{\hat{m}^2}{\mu^2}\right)^{3 \ep/2} \biggl\{ }
+\frac{\hat{\gamma}_{qg}^{(0)}}{6} \big(2 \hat{\gamma}_{gq}^{(1)} - \gamma_{gq}^{(1)}\big) + \delta m_1^{(-1)} \hat{\gamma}_{qg}^{(0)} \gamma_{gq}^{(0)}
\biggr]
\nonumber \\ && \phantom{\left(\frac{\hat{m}^2}{\mu^2}\right)^{3 \ep/2} \biggl\{ }
+\frac{1}{\ep} \biggl[
\frac{\hat{\gamma}_{qq}^{(2), \rm PS}}{3} - \frac{N_F}{3} \hat{\tilde{\gamma}}_{qq}^{(2), \rm PS} + \hat{\gamma}_{qg}^{(0)} a_{gq,Q}^{(2)} - \gamma_{gq}^{(0)} a_{Qg}^{(2)}
\nonumber \\ && \phantom{\left(\frac{\hat{m}^2}{\mu^2}\right)^{3 \ep/2} \biggl\{ }
-4 \big(\beta_0+\beta_{0,Q}\big) a_{Qq}^{(2), \rm PS} - \frac{\zeta_2}{16} \hat{\gamma}_{qg}^{(0)} \gamma_{gq}^{(0)} \big(\gamma_{gg}^{(0)} - \gamma_{qq}^{(0)} + 6 \beta_0\big)
\nonumber \\ && \phantom{\left(\frac{\hat{m}^2}{\mu^2}\right)^{3 \ep/2} \biggl\{ }
+\delta m_1^{(0)} \hat{\gamma}_{qg}^{(0)} \gamma_{gq}^{(0)} - \delta m_1^{(-1)} \hat{\gamma}_{qq}^{(1), \rm PS}
\biggr]
+a_{Qq}^{(3), \rm PS}
\Biggr\}.
\end{eqnarray}
Here $\gamma_{ij}^{(k)}$, with $k = 0,1,2$, denote the polarized anomalous dimensions 
\cite{Gross:1973ju,Georgi:1951sr,Sasaki:1975hk,Ahmed:1975tj,Altarelli:1977zs,Floratos:1977au,Curci:1980uw,
GonzalezArroyo:1979df,Moch:1999eb,Mertig:1995ny,SP_PS1,Moch:2014sna,Behring:2019tus},  $a_{ij}^{(k)}$, 
with $k = 1,2,3$,
is the constant part of the unrenormalized OME at $O(a_s^k)$, $\bar{a}_{ij}^{(k)}$, with $k = 1,2$,
denotes the $O(\ep)$ contribution of the unrenormalized OME at $O(a_s^k)$,
$\beta_k$ and $\beta_{Q,k}$ are the expansion coefficients of the QCD $\beta$-function in the 
$\overline{\rm MS}$--scheme and for massive contributions, $\delta m_k^{(l)}$ are the expansion coefficients
of the renormalized quark mass $m$, $\mu$ is the renormalization scale, $N_F$ denotes the number of light quark flavors,
and $\zeta_k = \sum_{l=1}^\infty (1/l^k)$, with $k \in \mathbb{N}, k \geq 2$, denotes the Riemann $\zeta$-function at integer
values. For details of the notation see Ref.~\cite{Bierenbaum:2009mv}. The two--loop results on $a_{ij}^{(k)}$ and 
$\bar{a}_{ij}^{(k)}$ are given in Ref.~\cite{POL19,Hasselhuhn:2013swa}, see also \cite{Buza:1996xr}.

Here and in the following we use the shorthand notations
\begin{eqnarray}
\hat{f}(x,N_F)   &\equiv& f(x,N_F+1) - f(x,N_F)\\
\tilde{f}(x,N_F) &\equiv& \frac{f(x,N_F)}{N_F}~.
\end{eqnarray} 

Renormalizing the mass in the on--shell scheme and the coupling constant in the ${\MS}$--scheme, we obtain the following expressions 
for
the renormalized pure singlet OME at two- and three--loop order 
\cite{Bierenbaum:2009mv},
\begin{eqnarray}
A_{Qq}^{(2),\rm PS, \MS}&=&
                -\frac{\hat{\gamma}_{qg}^{(0)}
                             \gamma_{gq}^{(0)}}{8}
                   \ln^2 \left(\frac{m^2}{\mu^2}\right)
                +\frac{\hat{\gamma}_{qq}^{(1), {\rm PS}}}{2}
                   \ln \left(\frac{m^2}{\mu^2}\right)
                +a_{Qq}^{(2),{\rm PS}}
                +\frac{\hat{\gamma}_{qg}^{(0)}
                             \gamma_{gq}^{(0)}}{8}\zeta_2~,
\label{AQq2PSMSren} 
\\
A_{Qq}^{(3),{\rm PS}, \MS}&=&
      \frac{\hat{\gamma}_{qg}^{(0)}\gamma_{gq}^{(0)}}{48}
                  \Bigl(
                         \gamma_{gg}^{(0)}
                        -\gamma_{qq}^{(0)}
                        +6\beta_0
                        +16\beta_{0,Q}
                  \Bigr)
              \ln^3 \left(\frac{m^2}{\mu^2}\right)
\NN\\ &&
  +              \biggl[
                         -\frac{\hat{\gamma}_{qq}^{(1),\rm PS}}{2}
                               \Bigl(
                                 \beta_0
                                +\beta_{0,Q}
                               \Bigr)
                        +\frac{\hat{\gamma}_{qg}^{(0)}}{8}
                               \Bigl(
                                 \hat{\gamma}_{gq}^{(1)}
                                -\gamma_{gq}^{(1)}
                               \Bigr)
                        -\frac{1}{8} \gamma_{gq}^{(0)}\hat{\gamma}_{qg}^{(1)}
                  \biggr]
              \ln^2 \left(\frac{m^2}{\mu^2}\right)
\NN\\ &&
  +   \biggl[
                         \frac{\hat{\gamma}_{qq}^{(2),{\rm PS}}}{2}
                        -\frac{N_F}{2} \hat{\tilde{\gamma}}_{qq}^{(2),{\rm PS}}
                        -2 a_{Qq}^{(2),{\rm PS}} (\beta_0+\beta_{0,Q})
                        +\frac{\hat{\gamma}_{qg}^{(0)}}{2} a_{gq,Q}^{(2)}
\NN\\ &&
                        -\frac{\gamma_{gq}^{(0)}}{2} a_{Qg}^{(2)}
                        -\frac{\zeta_2}{16} \hat{\gamma}_{qg}^{(0)} \gamma_{gq}^{(0)}
                          \Bigl(
                                 \gamma_{gg}^{(0)}
                                -\gamma_{qq}^{(0)}
                                +6\beta_0
                                +8\beta_{0,Q}
                          \Bigr)
                  \biggr]
              \ln \left(\frac{m^2}{\mu^2}\right)
\NN\\ &&
    +4(\beta_0+\beta_{0,Q})\overline{a}_{Qq}^{(2),{\rm PS}}
    +\gamma_{gq}^{(0)}\overline{a}_{Qg}^{(2)}
    -\hat{\gamma}_{qg}^{(0)}\overline{a}_{gq,Q}^{(2)}
\NN\\ &&
    +\frac{\zeta_3}{48} \gamma_{gq}^{(0)} \hat{\gamma}_{qg}^{(0)}
                  \Bigl(
                         \gamma_{gg}^{(0)}
                        -\gamma_{qq}^{(0)}
                        +6\beta_0
                  \Bigr)
    +\frac{\zeta_2}{16} \hat{\gamma}_{qg}^{(0)} \gamma_{gq}^{(1)}
\NN\\ &&
    -\delta m_1^{(1)} \hat{\gamma}_{qg}^{(0)} \gamma_{gq}^{(0)}
    +\delta m_1^{(0)} \hat{\gamma}_{qq}^{(1),{\rm PS}}
    +2 \delta m_1^{(-1)} a_{Qq}^{(2),{\rm PS}} 
    + {a_{Qq}^{(3),{\rm PS}}}~.     
\label{AQq3PSMSren} 
\end{eqnarray}

\begin{figure}[H]
\begin{minipage}[c]{0.18\linewidth}
     \includegraphics[width=1\textwidth]{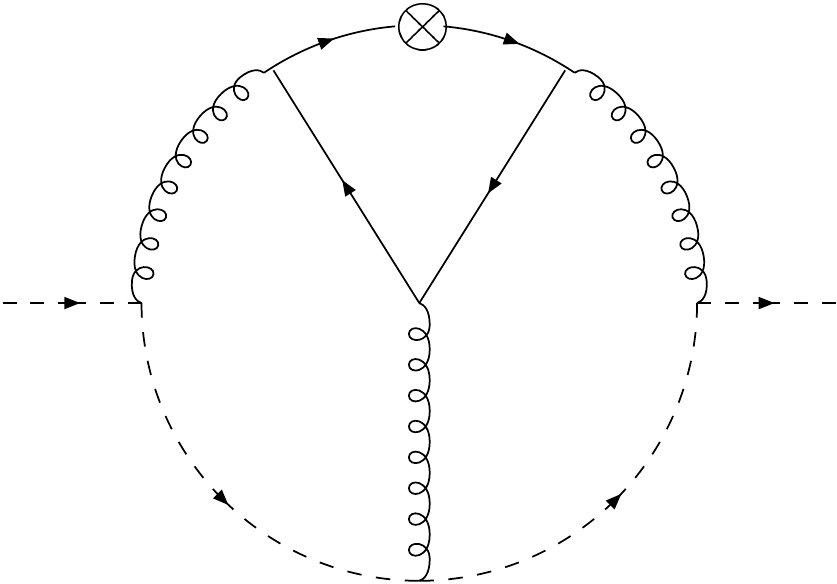}
\end{minipage}
\hspace*{0.5mm}
\begin{minipage}[c]{0.18\linewidth}
     \includegraphics[width=1\textwidth]{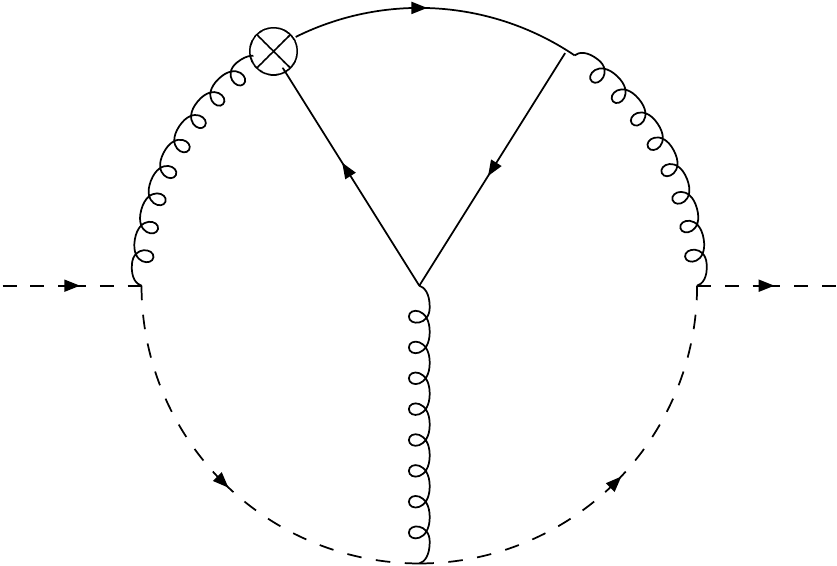}
\end{minipage}
\hspace*{0.5mm}
\begin{minipage}[c]{0.18\linewidth}
     \includegraphics[width=1\textwidth]{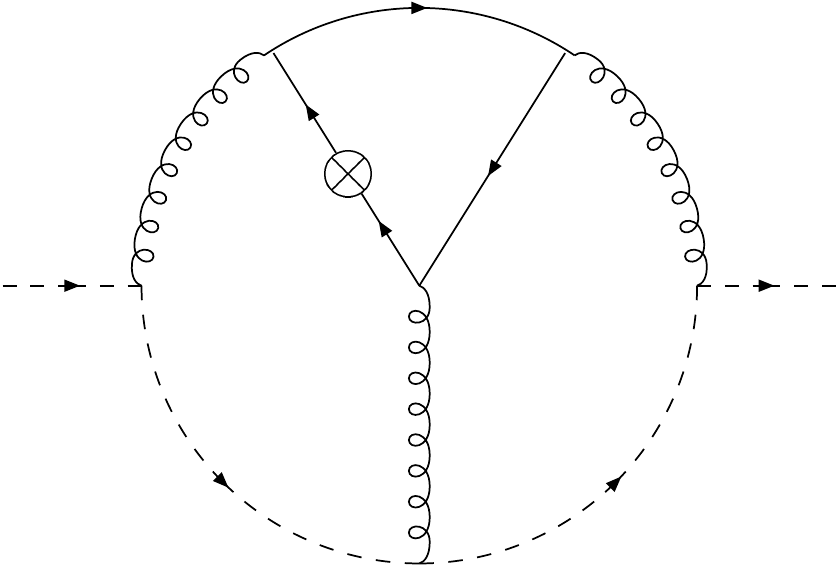}
\end{minipage}
\hspace*{0.5mm}
\begin{minipage}[c]{0.18\linewidth}
     \includegraphics[width=1\textwidth]{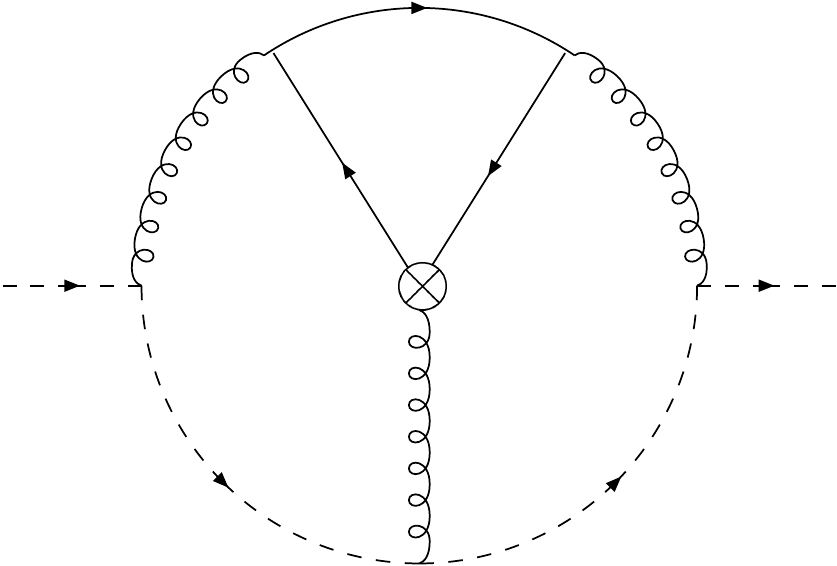}
\end{minipage}
\hspace*{0.5mm}
\begin{minipage}[c]{0.18\linewidth}
     \includegraphics[width=1\textwidth]{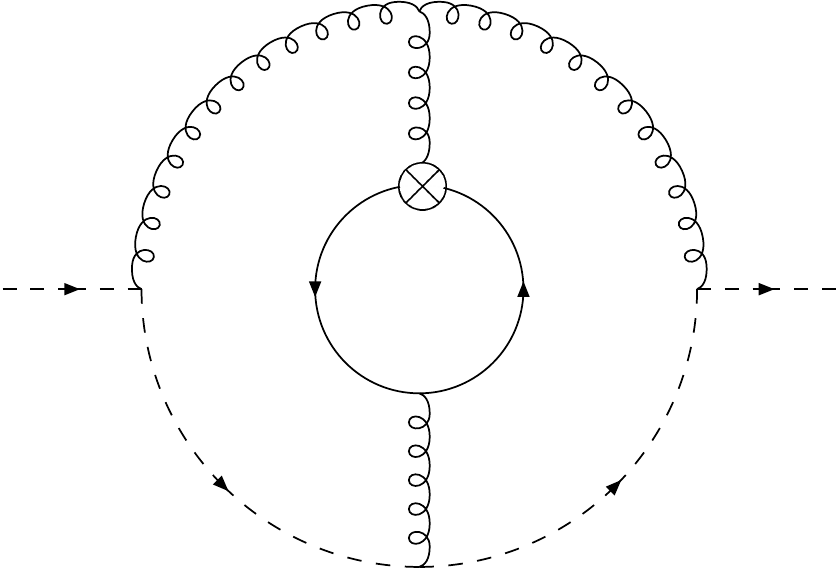}
\end{minipage}

\vspace*{2mm}
\begin{minipage}[c]{0.18\linewidth}
     \includegraphics[width=1\textwidth]{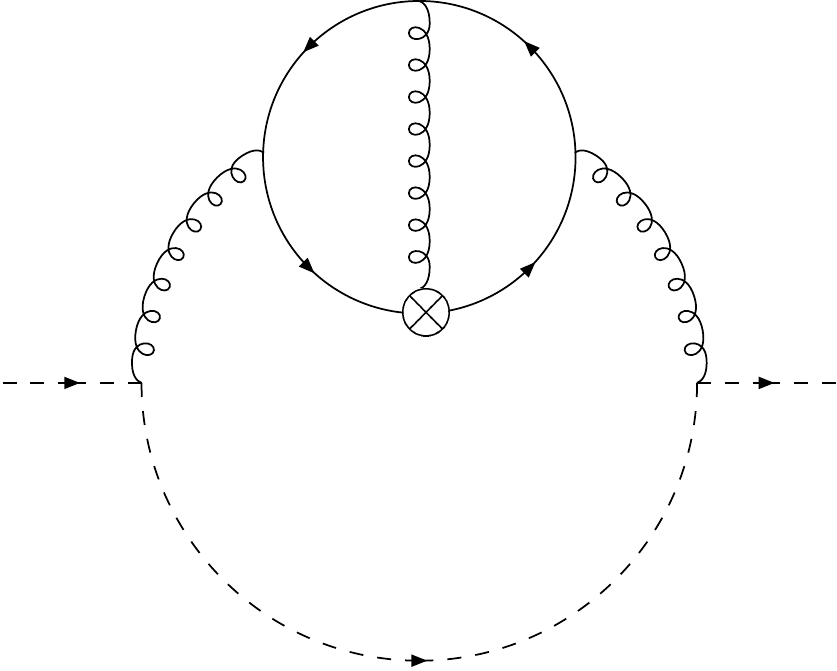}
\end{minipage}
\hspace*{0.5mm}
\begin{minipage}[c]{0.18\linewidth}
     \includegraphics[width=1\textwidth]{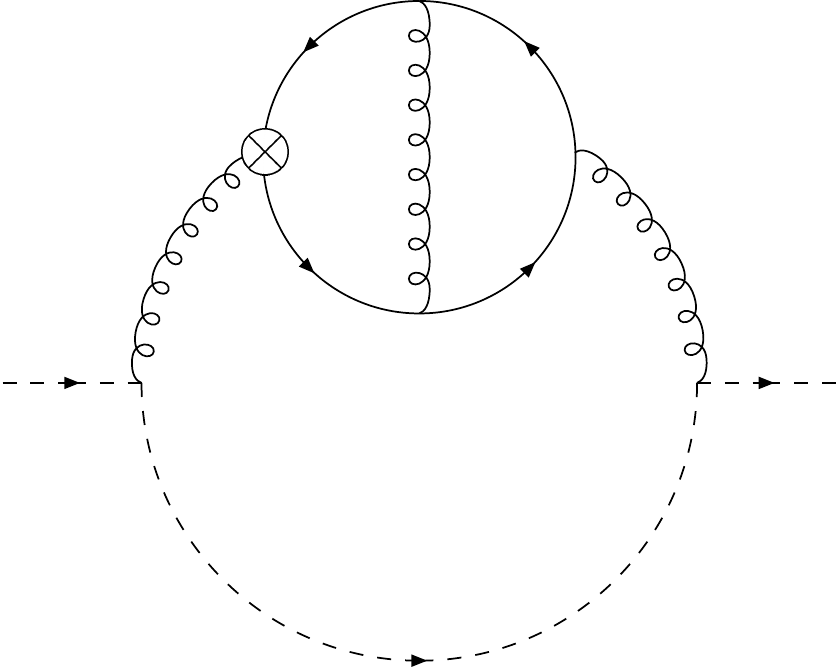}
\end{minipage}
\hspace*{0.5mm}
\begin{minipage}[c]{0.18\linewidth}
     \includegraphics[width=1\textwidth]{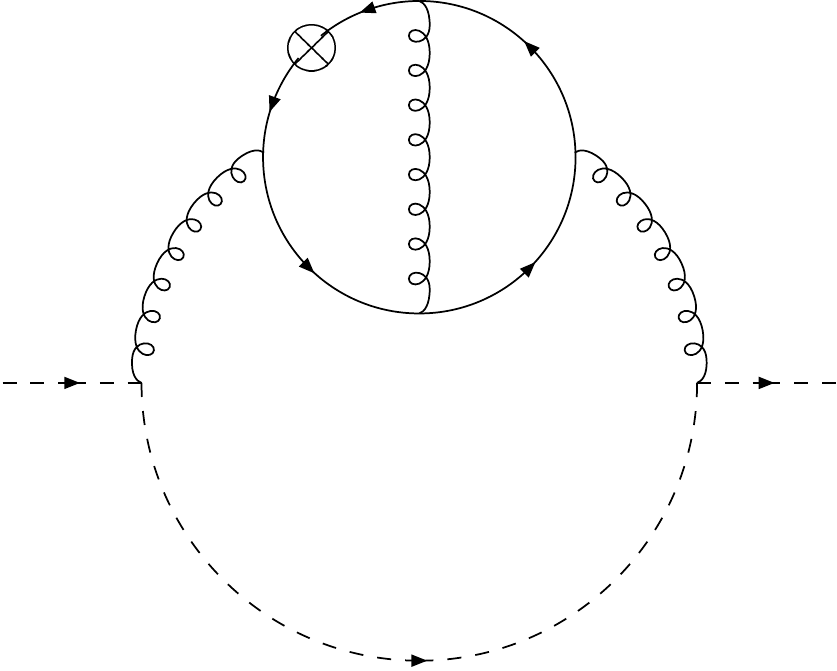}
\end{minipage}
\hspace*{0.5mm}
\begin{minipage}[c]{0.18\linewidth}
     \includegraphics[width=1\textwidth]{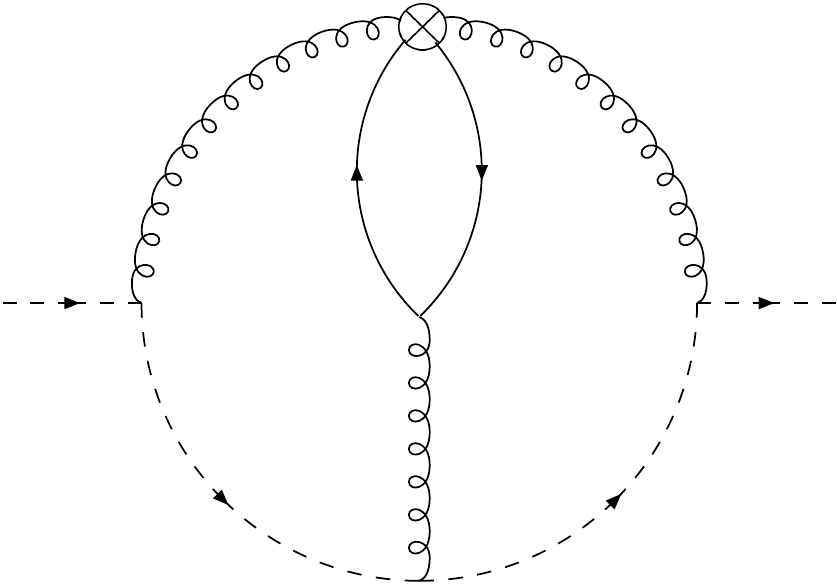}
\end{minipage}
\hspace*{0.5mm}
\begin{minipage}[c]{0.18\linewidth}
     \includegraphics[width=1\textwidth]{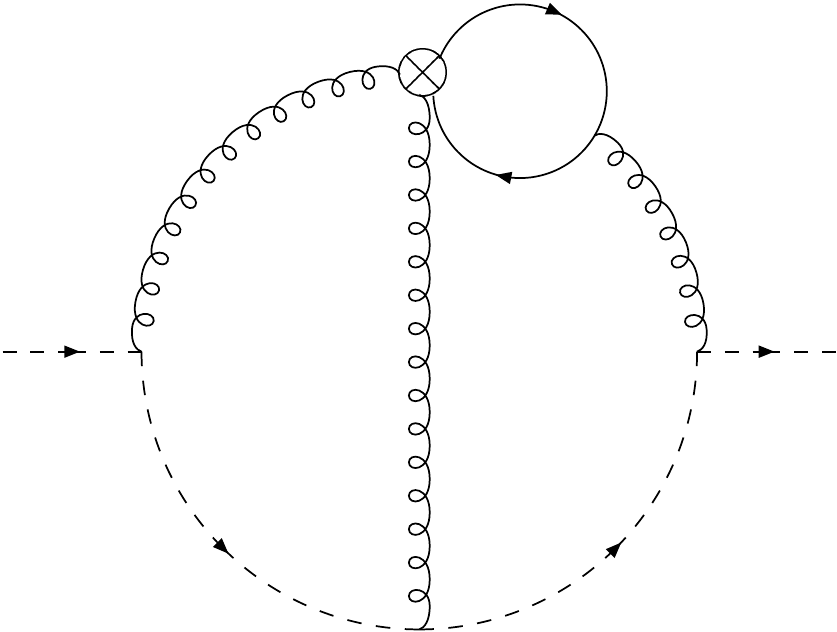} 
\end{minipage}

\vspace*{2mm}
\begin{minipage}[c]{0.18\linewidth}
     \includegraphics[width=1\textwidth]{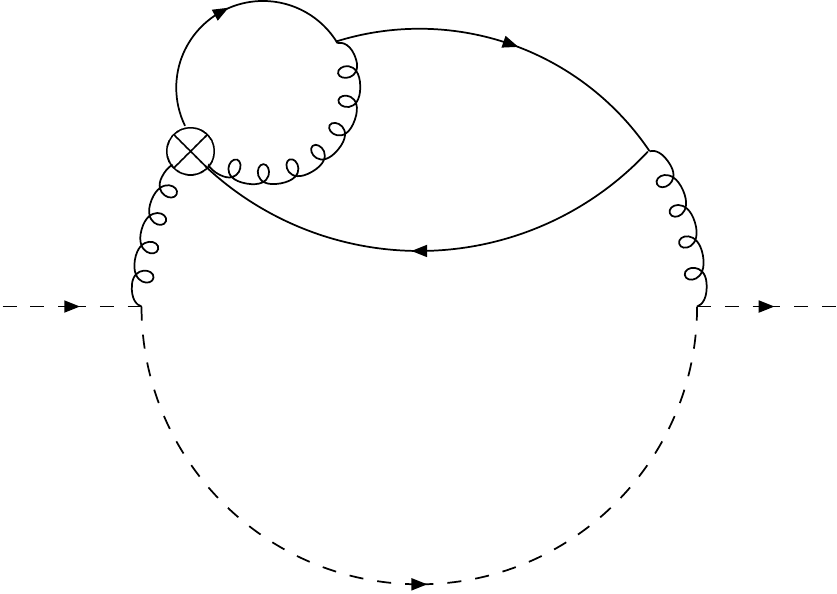} 
\end{minipage}
\hspace*{0.5mm}
\begin{minipage}[c]{0.18\linewidth}
     \includegraphics[width=1\textwidth]{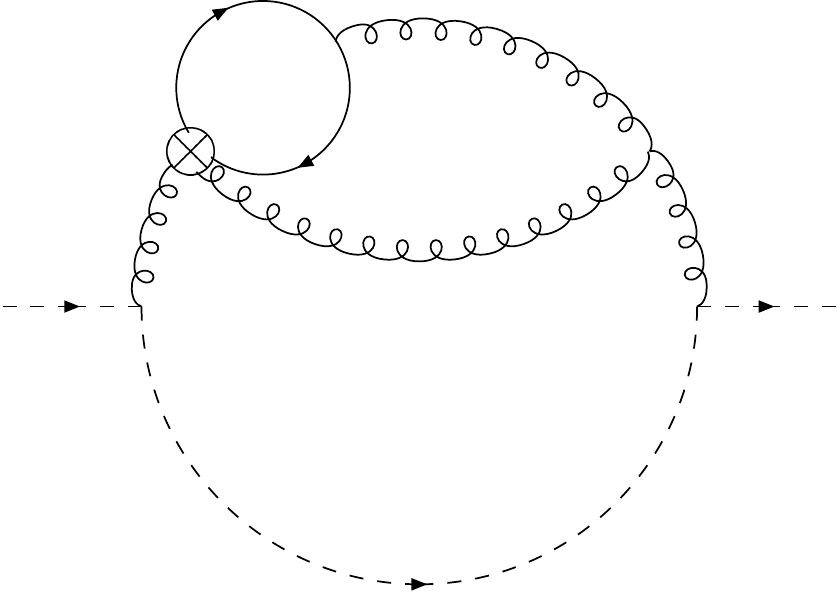} 
\end{minipage}
\hspace*{0.5mm}
\begin{minipage}[c]{0.18\linewidth}
     \includegraphics[width=1\textwidth]{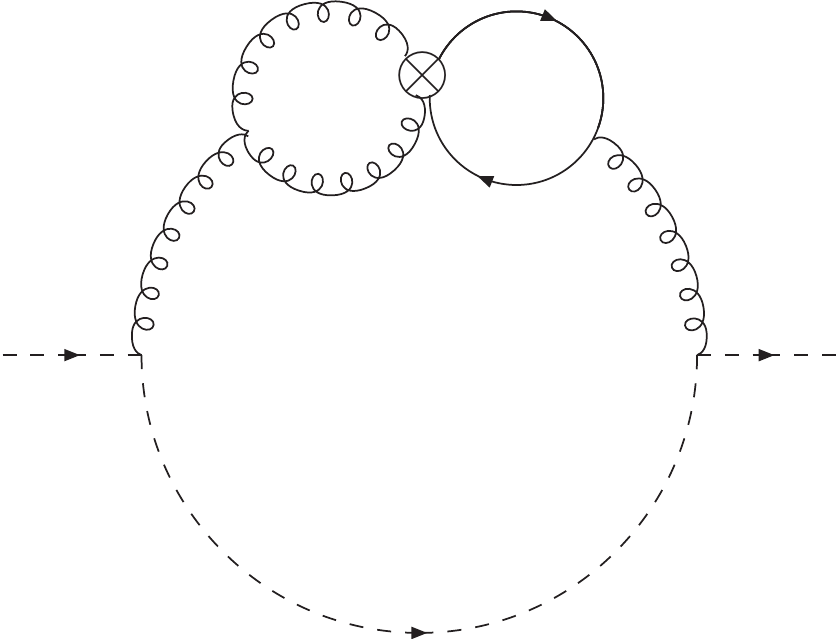}
\end{minipage}
\hspace*{0.5mm}
\begin{minipage}[c]{0.18\linewidth}
     \includegraphics[width=1\textwidth]{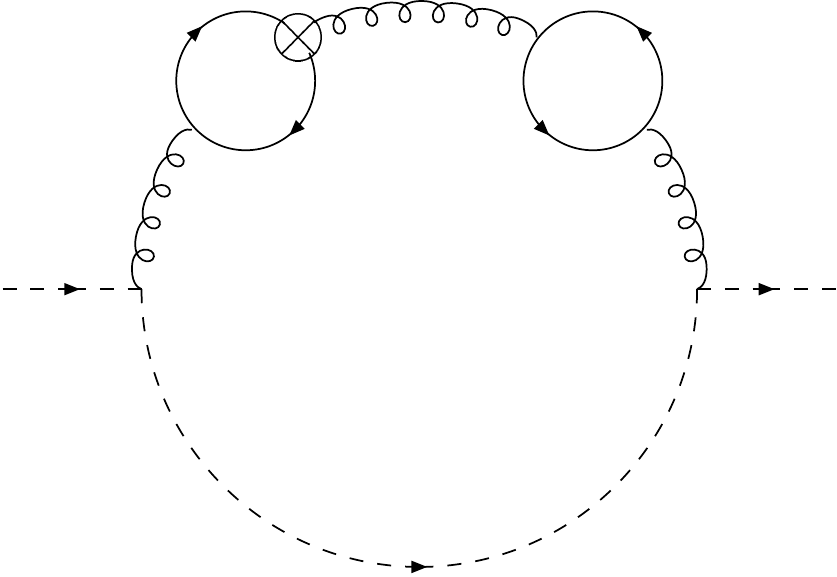}
\end{minipage}
\hspace*{0.5mm}
\begin{minipage}[c]{0.18\linewidth}
     \includegraphics[width=1\textwidth]{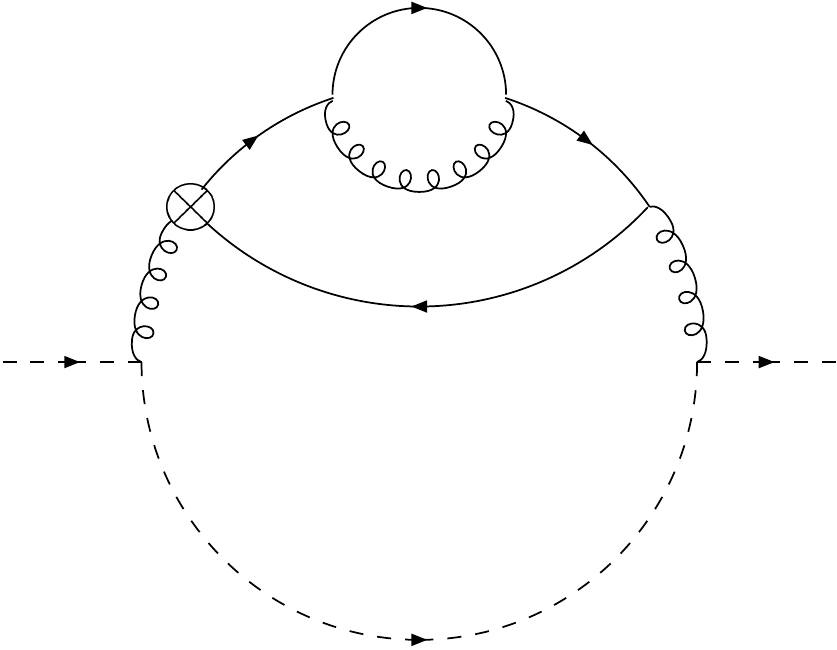}
\end{minipage}

\vspace*{2mm}
\begin{minipage}[c]{0.18\linewidth}
     \includegraphics[width=1\textwidth]{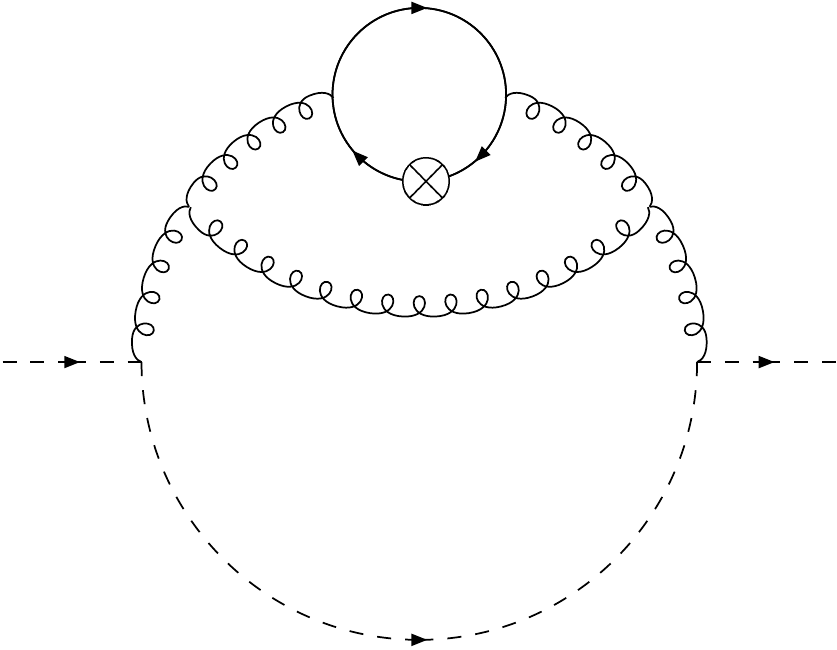}
\end{minipage}
\hspace*{0.5mm}
\begin{minipage}[c]{0.18\linewidth}
     \includegraphics[width=1\textwidth]{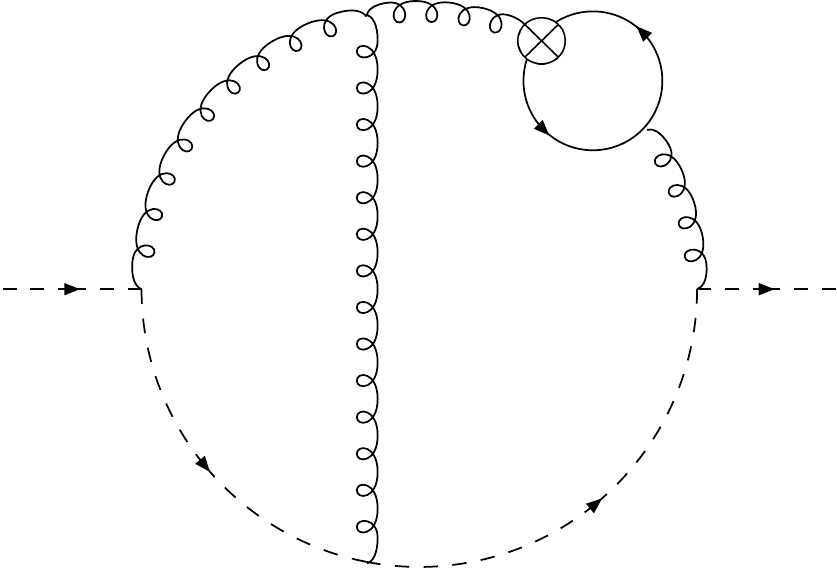}
\end{minipage}
\hspace*{0.5mm}
\begin{minipage}[c]{0.18\linewidth}
     \includegraphics[width=1\textwidth]{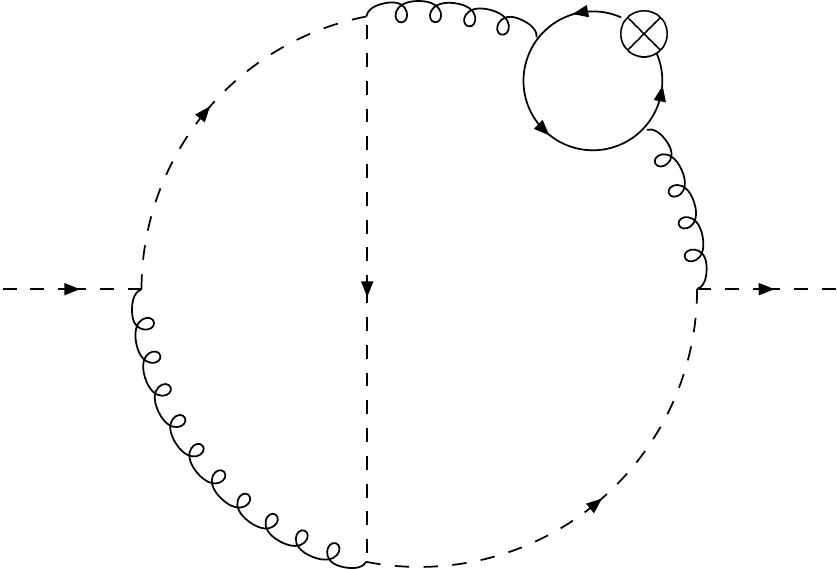}
\end{minipage}
\hspace*{0.5mm}
\begin{minipage}[c]{0.18\linewidth}
     \includegraphics[width=1\textwidth]{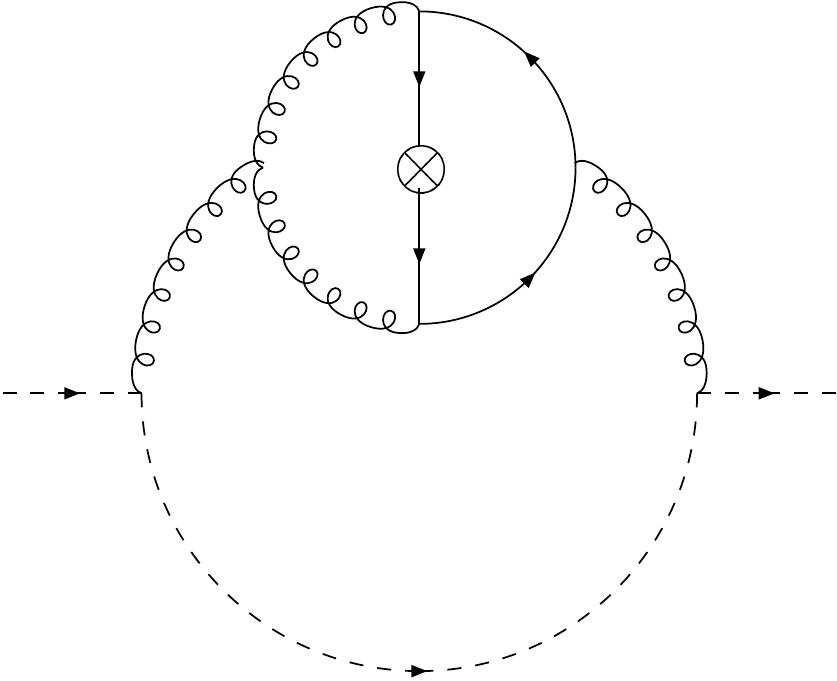}
\end{minipage}
\hspace*{0.5mm}
\begin{minipage}[c]{0.18\linewidth}
     \includegraphics[width=1\textwidth]{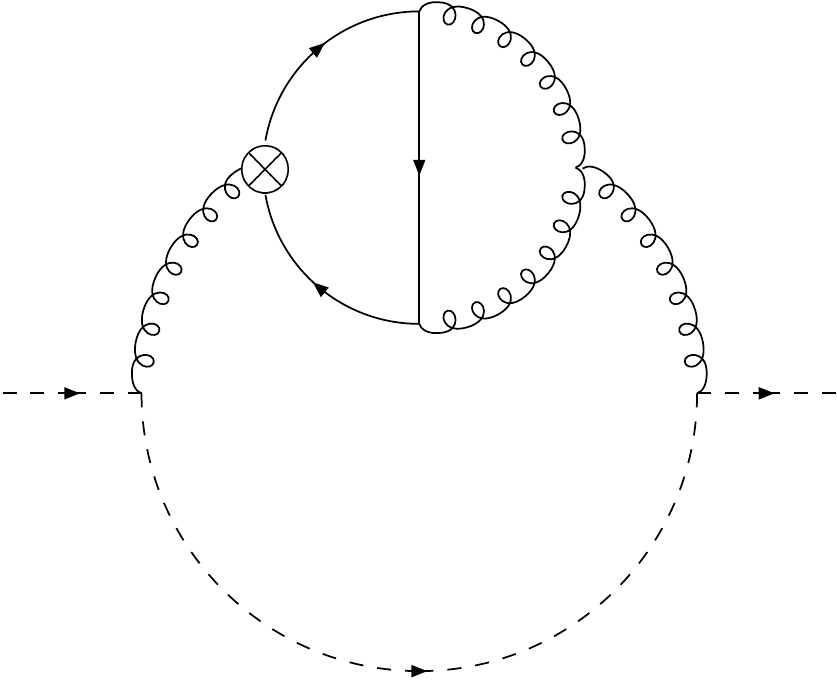}
\end{minipage}
\caption{\sf \small Sample of diagrams for $A_{Qq}^{(3), \rm PS}$. The dashed arrow lines represent
massless quarks, while the solid arrow lines represent massive quarks, and curly lines are gluons. 
The symbol $\otimes$ denotes the local operator insertion, see Ref.~\cite{Bierenbaum:2009mv}.
}
\label{samplediagrams}
\end{figure}

The connection of these OMEs to the massive Wilson coefficient in the asymptotic region has been described in
Ref.~\cite{Bierenbaum:2009mv}, Eq.~(2.14).
The polarized two--loop result was given in \cite{POL19,Buza:1996xr}. In this paper we present the three--loop result. In 
particular, we 
calculate the
constant part, $a_{Qq}^{(3),{\rm PS}}$, of the three--loop unrenormalized pure singlet polarized OME. The calculation in the 
polarized case is 
closely related to the unpolarized one \cite{Ablinger:2014nga}, since many of the required steps are common to or similar in both cases. 
The massive OME $A_{Qq}^{(3), \rm PS}$ consists of 125 Feynman diagrams, which we generated using {\tt QGRAF} \cite{Nogueira:1991ex}. A 
sample of the diagrams is shown in Figure~\ref{samplediagrams}. 
The Feynman rules for the local operator vertices can be found in 
Ref.~\cite{Bierenbaum:2009mv,Behring:2019tus}. The main difference between the polarized rules and the unpolarized ones is the presence 
of an additional factor of $\gamma_5$ in the former case, which requires the choice of a prescription in dimensional regularization.  
We performed the calculation using the Larin scheme \cite{Larin:1993tq}\footnote{For other schemes see Ref.~\cite{HVBM}. For 
a discussion of the necessary finite renormalizations see \cite{FINREN}.}, where $\gamma_5$ is expressed as
\begin{eqnarray}
\gamma^5 &=& \frac{i}{24} \ep_{\mu \nu \rho \sigma} \gamma^{\mu} \gamma^{\nu} \gamma^{\rho} \gamma^{\sigma}, \\
\slashed{\Delta} \gamma^5 &=& \frac{i}{6} \ep_{\mu \nu \rho \sigma} \Delta^{\mu} \gamma^{\nu} \gamma^{\rho} \gamma^{\sigma},
\end{eqnarray}
after which the Levi-Civita symbols can be contracted in $D$ dimensions using
\begin{eqnarray}
\ep_{\mu \nu \rho \sigma} \ep^{\alpha \lambda \tau \gamma} = -{\rm Det}\left[g_{\omega}^{\beta}\right], \quad\,\,
\beta = \alpha, \lambda, \tau, \gamma; \,\,\,\, \omega = \mu, \nu, \rho, \sigma.
\end{eqnarray}
In general, the calculation of OMEs requires the projection of the corresponding Green functions $\hat{G}^{ij}_l$, which is straightforward in the case of gluonic OMEs, but
is more subtle in the case of quarkonic polarized OMEs. In Ref.~\cite{Behring:2019tus}, we showed that the correct projector in the Larin scheme is given by
\begin{equation}
P_q \hat{G}_l^{ij} =  -\delta_{ij} \frac{i \left(\Delta.p\right)^{-N-1}}{4 N_c (D-2) (D-3)} \ep_{\mu \nu p \Delta}
{\rm tr} \left[\slashed{p} \gamma^{\mu} \gamma^{\nu} \hat{G}_l^{ij}\right],
\label{projector}
\end{equation}
where $N_c$ is the number of QCD colors, $p$ is the momentum of the external massless quark, $N$ is the Mellin variable, and $\Delta$ is a $D$-dimensional light-like vector.
Using this projector, we were able to extract the corresponding anomalous dimensions from the poles of the OMEs in \cite{Behring:2019tus}, and after performing a finite renormalization,
we found the results to agree with those presented in \cite{Moch:2014sna} in the so called M-scheme \cite{Matiounine:1998re}.
The full finite renormalization required to transform also the
constant term $a_{Qq}^{(3),{\rm PS}}$ to the M-scheme is at present unknown, so in this paper we stick to the Larin scheme.

The propagators, vertices and operator insertions from the output of {\tt QGRAF} were replaced by the corresponding Feynman rules using a {\tt FORM} \cite{Tentyukov:2007mu} program \cite{Bierenbaum:2009mv}, 
which also allowed us to introduce the projector (\ref{projector}) and to perform the Dirac algebra in the numerator of the Feynman 
integrals. After this, we ended up with a linear combination of a large number of scalar
integrals,
\begin{equation}
A_{Qq}^{(3),\rm PS}(N) = \sum_i c_i J_i(N)~,
\end{equation}
where the functions $J_i(N)$ are the scalar integrals and the $c_i$'s are factors containing scalar products not involving the loop 
momenta.
As we have done in the past for the calculation of unpolarized OMEs, we multiplied the result by an auxiliary variable $t$ raised to the power of the Mellin moment $N$, and summed up to $N=\infty$, that is,
we computed
\begin{equation}
\sum_{N=0}^{\infty} t^N \sum_i c_i J_i(N)~.
\end{equation}
This allowed us to rewrite all operator insertions in terms of artificial propagators, after which 
it became possible to reduce the scalar integrals to master integrals using integration
by parts identities. For this we used the {\tt C++} package {\tt Reduze 2} \cite{REDUZE}\footnote{
The package {\tt Reduze 2} uses the packages {\tt FERMAT} \cite{FERMAT} and {\tt Ginac}
\cite{Bauer:2000cp}.}. We ended up with a linear combination of 
master integrals,
\begin{equation}
A_{Qq}^{(3),\rm PS}(t) =
\sum_{N=0}^{\infty} t^N A_{Qq}^{(3),\rm PS}(N) = 
\sum_i r_i(t,D) \sum_{N=0}^{\infty} t^N M_i(N),
\end{equation}
where the functions $M_i(N)$ are the master integrals and $r_i(t,D)$ are rational functions in $t$ and the dimension $D$.
What remained was the calculation of the master integrals and the extraction of the $N$th coefficient of the expansion
in $t$ of $A_{Qq}^{(3),\rm PS}(t)$ as a function of $N$. 
The master integrals turn out to be the same ones needed in the unpolarized case. Details on their calculation can be found in 
Ref.~\cite{Ablinger:2014nga}. In their computation we use difference field and ring techniques as implemented in the packages {\tt 
Sigma}
\cite{SIG1,SIG2}, {\tt EvaluateMultiSums} and {\tt SumProduction} \cite{EMSSP}, which also make use of the package {\tt HarmonicSums} 
\cite{Vermaseren:1998uu,Blumlein:1998if,HARMONICSUMS,Ablinger:PhDThesis,Ablinger:2011te,Ablinger:2013cf,
Ablinger:2014bra}.\footnote{For a recent survey on the different calculation techniques see Ref.~\cite{Blumlein:2018cms}.}
We have checked our results comparing the moments for fixed values of $N$ with corresponding results obtained by using {\tt MATAD}
\cite{Steinhauser:2000ry} for $N = 3, 5, 7$.

\section{Results}
\label{sec:3}

\vspace*{1mm}
\noindent
We now present the result for the $O(\ep^0)$ term of the unrenormalized pure singlet polarized OME, namely, $a_{Qq}^{(3), \rm 
PS}$. In 
$N$ space, this term 
is given by harmonic sums~\cite{Vermaseren:1998uu,Blumlein:1998if}
\begin{eqnarray}
S_{i_n, i_{n-1}, \cdots, i_2, i_1}(N) = \sum_{k=1}^N \frac{{\rm sign}(i_n)^k}{k^{i_n}} S_{i_{n-1}, \cdots i_2, 
i_1}(k),~~S_\emptyset 
= 1,
\end{eqnarray}
and generalized harmonic sums, cf.~\cite{Ablinger:2013cf}, at rational weights $a_i \in \mathbb{Q}$
\begin{eqnarray}
S_{i_n i_{n-1} \cdots i_2 i_1}\left(a_n, a_{n-1}, \cdots, a_2, a_1; N\right) = \sum_{k=1}^N \frac{a_n^k}{k^{i_n}} S_{i_{n-1}, \cdots i_2, i_1}\left(a_{n-1}, \cdots, a_2, a_1; k\right),
\end{eqnarray}
with $i_k \in \mathbb{N} \backslash \{0\}$. In the following, 
we will use the shorthand notation
\begin{equation}
S_{i_n, \cdots, i_1} \equiv S_{i_n, \cdots, i_1}(N) \quad~~\text{and}~~\quad S_{i_n, \cdots, i_1}\left(a_n \cdots, a_1\right) 
\equiv S_{i_n, \cdots, i_1}\left(a_n \cdots, a_1; N\right).
\end{equation}
One obtains

\begin{eqnarray}
\label{eqaQq3N}
\lefteqn{a_{Qq}^{(3), \rm PS}(N) =} \nonumber\\ &&
\textcolor{blue}{C_F T_F \big(C_A -2 C_F \big)} \Biggl\{
         \frac{2^{4-N} P_2}{N^3 (N+1)^2} \big[
                 -S_{2,1}({{2,1}})
                 +S_{1,1,1}({{2,1,1}})
                 +7 \zeta_3
         \big]
\nonumber \\ &&
         +\frac{32}{N^3} (5 N-2) \biggl\{
                 \frac{2^{-N}}{(N-1) (N+1)^2} \big[
                         -S_2 S_1({{2}})
                         +S_{2,1}({{1,2}})
                 \big]
                 +\frac{2^N}{N} \biggl[
                          S_{1,1}\biggl({{\frac{1}{2},1}}\biggr)
\nonumber \\ &&
                         +S_{1,1}\biggl({{1,\frac{1}{2}}}\biggr)
                         -S_1 S_1\biggl({{\frac{1}{2}}}\biggr)
                         -S_2\biggl({{\frac{1}{2}}}\biggr)
                 \biggr]
                 +2^N \biggl[
                         \frac{1}{2} S_1^2 S_1\biggl({{\frac{1}{2}}}\biggr)
                         +\frac{1}{2} S_2 S_1\biggl({{\frac{1}{2}}}\biggr)
\nonumber \\ &&
                         +S_1 S_2\biggl({{\frac{1}{2}}}\biggr)
                         +S_3\biggl({{\frac{1}{2}}}\biggr)
                         -S_{1,1,1}\biggl({{\frac{1}{2},1,1}}\biggr)
                         -S_{1,1,1}\biggl({{1,\frac{1}{2},1}}\biggr)
                         -S_{1,1,1}\biggl({{1,1,\frac{1}{2}}}\biggr)
                 \biggr]
         \biggr\}
\nonumber \\ &&
         +\frac{2^{4-N} \big(N^2+N+2\big)}{(N-1) N (N+1)}
 \big[ S_{1,2}({{2,1}})
      -S_3({{2}})
         \big]
         +\frac{2^{5-N} (2 N-1)}{(N-1) N (N+1)^2} \big[
                  S_{2,1}({{1,2}})
                 -S_2 S_1({{2}})
         \big]
\nonumber \\ &&
         +32 F_1 \biggl\{
                 \biggl[
                          S_{1,1}\biggl({{2,\frac{1}{2}}}\biggr)
                         -S_1\biggl({{\frac{1}{2}}}\biggr) S_1({{2}})
                 \biggr] S_2
                 -S_1({{2}}) S_3\biggl({{\frac{1}{2}}}\biggr)
                 -S_2\biggl({{\frac{1}{2}}}\biggr) S_{1,1}({{2,1}})
\nonumber \\ &&
                 +\big[
                          S_{2,1}({{1,2}})
                         +S_{2,1}({{2,1}})
                         -S_{1,1,1}({{2,1,1}})
                         -7 \zeta_3
                 \big] S_1\biggl({{\frac{1}{2}}}\biggr)
                 +S_{2,2}\biggr({{2,\frac{1}{2}}}\biggl)
                 +S_{3,1}\biggr({{\frac{1}{2},2}}\biggl)
\nonumber \\ &&
                 +S_{2,1,1}\biggr({{\frac{1}{2},2,1}}\biggl)
                 -S_{2,1,1}\biggr({{1,2,\frac{1}{2}}}\biggl)
                 -S_{2,1,1}\biggr({{2,\frac{1}{2},1}}\biggl)
                 -S_{2,1,1}\biggr({{2,1,\frac{1}{2}}}\biggl)
\nonumber \\ &&
                 +S_{1,1,1,1}\biggr({{2,\frac{1}{2},1,1}}\biggl)
                 +S_{1,1,1,1}\biggr({{2,1,\frac{1}{2},1}}\biggl)
                 +S_{1,1,1,1}\biggr({{2,1,1,\frac{1}{2}}}\biggl)
                 -\frac{B_4}{2}
         \biggr\}
 \Biggr\}
\nonumber \\ &&
 +\textcolor{blue}{C_F T_F^2 N_F} \Biggl\{
         -\frac{32 (N+2) P_{22}}{243 N^5 (N+1)^5}
         -\frac{(N+2)}{N^3 (N+1)^3} \biggl[
                 \frac{16}{27} S_1^2
                 +\frac{208}{27} S_2
                 +\frac{16}{9} \zeta_2
         \biggr] P_1
\nonumber \\ &&
         +\frac{16}{3} F_1 \biggl[
                 \biggl(
                         \frac{13}{3} S_2
                         +\zeta_2
                 \biggr) S_1
                 +\frac{1}{9} S_1^3
                 +\frac{110}{9} S_3
                 -\frac{14}{3} \zeta_3
         \biggr]
         +\frac{32 (N+2) P_9 S_1}{81 N^4 (N+1)^4}
 \Biggr\}
\nonumber \\ &&
 +\textcolor{blue}{C_F T_F^2} \Biggl\{
         \frac{32}{3} F_1 \biggl[
                 \biggl(
                         \frac{5}{3} S_2
                         -\zeta_2
                 \biggr) S_1
                 -\frac{1}{9} S_1^3
                 +\frac{16}{9} S_3
                 -4 S_{2,1}
                 +\frac{32 \zeta_3}{3}
         \biggr]
         +\frac{32}{9} \zeta_2 F_2
\nonumber \\ &&
+\frac{32}{27 N^3 (N+3) (N+4)} \biggl[
          \frac{P_{11} S_1^2}{(N+1)^3}
         -\frac{P_{10} S_2}{(N+1)^2}
         +\frac{2 P_{27}-6 N P_{25} S_1}{9 N^2 (N+1)^4}
\biggr]
\Biggr\}
\nonumber \\ &&
 +\textcolor{blue}{C_A C_F T_F} \Biggl\{
        \frac{4 P_{20} S_1^2 + 4 P_{23} S_2}{27 N^4 (N+1)^4 (N+2)}
         +\frac{8 P_{30}}{243 (N-1) N^6 (N+1)^6 (N+2)}
\nonumber \\ &&
         +32 F_1 \biggl[
                 \frac{9}{2} \zeta_4
                 -S_1^2 \biggl(
                         \frac{17}{24} S_2
                         +\frac{\zeta_2}{8}
                 \biggr)
                 -\frac{S_1^4}{144}
                 -\biggl(
                          \frac{7}{6} S_1^2
                         +\frac{3}{4} \zeta_2
                 \biggr) S_{-2}
                 +\frac{5}{6} S_{2,1,1}
                 -\frac{3}{8} \zeta_2 S_2
         \biggr]
\nonumber \\ &&
         +\biggl[
                 -\frac{8 P_{28}}{81 N^5 (N+1)^5 (N+2)}
                 -\frac{4 P_7 S_2 + 12 \zeta_2 P_4}{9 N^3 (N+1)^3}
                 -\frac{8 \big(137 N^2+137 N-334\big)}{9 N^2 (N+1)^2} S_3
\nonumber \\ &&
                 +\frac{16 \big(35 N^2+35 N-18\big)}{3 N^2 (N+1)^2} S_{-2,1}
                 +\frac{8 \big(11 N^2+11 N-10\big) \zeta_3}{3 N^2 (N+1)^2}
         \biggr] S_1
         -\frac{4 P_4 S_1^3}{27 N^3 (N+1)^3}
\nonumber \\ &&
         -\frac{2 \big(29 N^2+29 N-74\big)}{3 N^2 (N+1)^2} S_2^2
         -\frac{8 P_{18} S_3}{27 N^3 (N+1)^3}
         -\frac{4 \big(167 N^2+167 N-358\big)}{3 N^2 (N+1)^2} S_4
\nonumber \\ &&
         +\biggl[
                 \frac{32 (N+2)}{3 N^3 (N+1)^3} \big(N^3-9 N^2+16 N+4\big) S_1
                 +\frac{16 P_{24}}{3 (N-1) N^4 (N+1)^4 (N+2)}
\nonumber \\ &&
                 -\frac{64 \big(7 N^2+7 N-13\big)}{3 N^2 (N+1)^2} S_2
         \biggr] S_{-2}
         +\frac{16 \big(3 N^2+3 N+2\big)}{3 N^2 (N+1)^2} S_{-2}^2
         +\frac{8 \zeta_3 P_{14} - 48 P_{13} S_{-2,1}}{9 N^3 (N+1)^3}
\nonumber \\ &&
         -\biggl[
                  \frac{8 P_{12}}{3 N^3 (N+1)^3}
                 +\frac{8 \big(69 N^2+69 N-94\big)}{3 N^2 (N+1)^2} S_1
         \biggr] S_{-3}
         -\frac{16 \big(31 N^2+31 N-50\big)}{3 N^2 (N+1)^2} S_{-4}
\nonumber \\ &&
         +\frac{8 (N-1) P_3 S_{2,1}}{3 N^3 (N+1)^3}
         +\frac{24 (N-2) (N+3)}{N^2 (N+1)^2} S_{3,1}
         +\frac{64 \big(3 N^2+3 N-2\big)}{N^2 (N+1)^2} S_{-2,2}
\nonumber \\ &&
         +\frac{32 \big(23 N^2+23 N-22\big)}{3 N^2 (N+1)^2} S_{-3,1}
         -\frac{64 \big(13 N^2+13 N-2\big)}{3 N^2 (N+1)^2} S_{-2,1,1}
         +\frac{4 P_{19} \zeta_2}{9 N^4 (N+1)^4}
 \Biggr\}
\nonumber \\ &&
 +\textcolor{blue}{C_F^2 T_F} \Biggl\{
         -\frac{4 P_{29}}{3 N^6 (N+1)^6 (N+2)}
         +\frac{4 P_{17} S_1^2 - 4 P_{21} S_2}{3 N^4 (N+1)^4 (N+2)}
         +\frac{16 P_{15} S_3}{9 N^3 (N+1)^3}
\nonumber \\ &&
         +\biggl(
                 \frac{8 P_{26}}{3 N^5 (N+1)^5 (N+2)}
                 +\frac{4 P_6 S_2}{3 N^3 (N+1)^3}
         \biggr) S_1
         +64 F_1 \biggl[
                 -\frac{9}{4} \zeta_4
                 -\frac{P_5 \zeta_2}{32 N^2 (N+1)^2}
\nonumber \\ &&
                 +\frac{3 N^2-3 N-4}{16 N (N+1)}
 \biggl(\frac{1}{9} S_1^3
                         +\zeta_2 S_1
                 \biggr)
                 +\biggl(
                          S_{2,1}
                         -\frac{5}{36} S_3
                         -\frac{7}{12} \zeta_3
                 \biggr) S_1
                 +\biggl(
                         \frac{5}{48} S_2
                         +\frac{\zeta_2}{16}
                 \biggr) S_1^2
\nonumber \\ &&
                 +\frac{1}{288} S_1^4
                 -\frac{23}{96} S_2^2
                 +\frac{17}{48} S_4
                 +S_{3,1}
                 -\frac{13}{6} S_{2,1,1}
                 -\frac{3}{16} \zeta_2 S_2
         \biggr]
         -\frac{32 P_8 S_{2,1} + 4 P_{16} \zeta_3}{3 N^3 (N+1)^3}
 \Biggr\},
\end{eqnarray}
with the functions
\begin{eqnarray}
F_1 &=& \frac{(N-1) (N+2)}{N^2 (N+1)^2}~, \label{F1inN} \\
F_2 &=& \frac{(N-3) (N+2) (2 N+1)}{N^3 (N+1)^2}~. 
\label{F2inN}
\end{eqnarray}
The constants $B_4$ and $B_5$ are defined as
\begin{eqnarray}
B_4 &=& 16 {\rm Li}_4\left(\frac{1}{2}\right)
+\frac{2}{3} \ln^4(2)
-4 \ln^2(2) \zeta_2
-\frac{13}{2} \zeta_4 
\\
B_5 &=& 16 {\rm Li}_5\left(\frac{1}{2}\right)
+8 \ln(2) {\rm Li}_4\left(\frac{1}{2}\right)
+\frac{101}{48} \zeta_2 \zeta_3
-\frac{443}{32} \zeta_5
\nonumber \\ &&
+\frac{1}{5} \ln^5(2)
-\frac{2}{3} \ln^3(2) \zeta_2
-\frac{35}{4} \ln(2) \zeta_3
+\frac{61}{16} \ln(2) \zeta_4.
\end{eqnarray}
They are linear combinations of multiple zeta values \cite{Blumlein:2009cf} and $\Li_n(x) = \sum_{k=1}^\infty x^k/k^n, x \in [-1,1]$
denotes the classical polylogarithm.
The polynomials $P_i$ are given by
\begin{eqnarray}
  P_1 &=& 11 N^3-3 N^2+10 N+6, \\ 
  P_2 &=& N^4+3 N^3+2 N^2+6 N-4, \\
  P_3 &=& 6 N^4+38 N^3+52 N^2+81 N+42, \\ 
  P_4 &=& 11 N^4+22 N^3-23 N^2-70 N-12, \\ 
  P_5 &=& 35 N^4+64 N^3+28 N^2-13 N-6, \\ 
  P_6 &=& 71 N^4+8 N^3-121 N^2+66 N+72, \\ 
  P_7 &=& 203 N^4+394 N^3-125 N^2-928 N-192, \\ 
  P_8 &=& 6 N^5+21 N^4-26 N^3-48 N^2-39 N-22, \\ 
  P_9 &=& 58 N^5+25 N^4+167 N^3-94 N^2-96 N-36, \\ 
  P_{10} &=& 64 N^5+497 N^4+614 N^3-545 N^2-126 N-360, \\ 
  P_{11} &=& 2 N^6+15 N^5+179 N^4+471 N^3+503 N^2+774 N+360, \\ 
  P_{12} &=& 6 N^6+15 N^5-24 N^4+29 N^2-138 N-12, \\ 
  P_{13} &=& 6 N^6+15 N^5+24 N^4-8 N^3-3 N^2+78 N+20, \\ 
  P_{14} &=& 9 N^6+54 N^5-148 N^4-377 N^3-1421 N^2-679 N+330, \\ 
  P_{15} &=& 36 N^6+108 N^5-63 N^4-531 N^3-1001 N^2-623 N-14, \\ 
  P_{16} &=& 48 N^6+48 N^5-501 N^4-378 N^3-881 N^2-524 N+268, \\ 
  P_{17} &=& 117 N^6+566 N^5+1137 N^4+1348 N^3+944 N^2+208 N-48, \\ 
  P_{18} &=& 135 N^6+540 N^5+875 N^4+49 N^3-3686 N^2-3715 N+1086, \\ 
  P_{19} &=& 160 N^6+447 N^5+211 N^4+159 N^3+475 N^2-192 N-108, \\ 
  P_{20} &=& 79 N^7-52 N^6-1379 N^5-2938 N^4-1781 N^3+947 N^2+1308 N+468, \\ 
  P_{21} &=& 158 N^7+381 N^6-566 N^5-2683 N^4-3502 N^3-1860 N^2+312 N+416, \\ 
  P_{22} &=& 332 N^7+490 N^6+1167 N^5-1555 N^4+754 N^3+1140 N^2+792 N+216, \\ 
  P_{23} &=& 2125 N^7+8792 N^6+9505 N^5+632 N^4+3487 N^3+4535 N^2
\nonumber \\ && -7188 N-2556, \\ 
  P_{24} &=& 3 N^8+26 N^7+28 N^6-41 N^5+82 N^4+111 N^3-257 N^2-20 N-4, \\ 
  P_{25} &=& 13 N^8-146 N^7-1061 N^6-2606 N^5-2516 N^4+502 N^3+1746 N^2
\nonumber \\ && +1620 N+432, \\ 
  P_{26} &=& 12 N^9+32 N^8-5 N^7-194 N^6-493 N^5+76 N^4+1568 N^3+1596 N^2
\nonumber \\ && +704 N+112, \\ 
  P_{27} &=& 80 N^9-859 N^8-6334 N^7-16687 N^6-22150 N^5-15142 N^4-15840 N^3
\nonumber \\ && -15228 N^2-9720 N-2592, \\ 
  P_{28} &=& 968 N^9+5625 N^8+13824 N^7+19941 N^6+15627 N^5-5448 N^4-22490 N^3
\nonumber \\ && -12963 N^2-2772 N-108, \\ 
  P_{29} &=& 184 N^{11}+1168 N^{10}+3055 N^9+4058 N^8+2015 N^7-1489 N^6-2103 N^5
\nonumber \\ && +75 N^4+2179 N^3+2158 N^2+1060 N+216, \\ 
  P_{30} &=& 6472 N^{12}+35280 N^{11}+76634 N^{10}+67296 N^9-65784 N^8-151947 N^7
\nonumber \\ && +81392 N^6+112464 N^5-171758 N^4-51321 N^3-11844 N^2
\nonumber \\ && +18684 N+7776~.
\end{eqnarray}

The Mellin inversion of (\ref{eqaQq3N}) leads to generalized harmonic polylogarithms (HPLs) of argument $x$ 
\cite{Ablinger:2013cf}.
These can be transformed to standard harmonic polylogarithms \cite{Remiddi:1999ew}
over the alphabet $\{-1, 0, 1\}$ evaluated at different arguments, which
can be done with the help of the Mathematica package {\tt HarmonicSums}. The harmonic polylogarithms are given by
\begin{eqnarray}
H_{b,\vec{a}}(x) = \int_0^x dy f_b(y) H_{\vec{a}}(y),~~~~a_i, b \in \{-1,0,1\},~~~~H_\emptyset = 1,
\end{eqnarray}
with $f_0 = 1/y, f_{-1} = 1/(1+y), f_1 = 1/(1-y)$.

In the case of $a_{Qq}^{(3), \rm PS}$, we obtain the usual harmonic polylogarithms at argument $x$
and a set of harmonic polylogarithms at argument $(1 - 2x)$.
This representation is of advantage for later numerical representations\footnote{Note that using the 
harmonic polylogarithms at a different continuous argument implies in general a new class of functions with only exceptional 
relations.}. The presence of the argument $(1-2x)$ 
in the OME will require a modification of the Mellin convolution. In intermediate steps we observed the supports $[0,1/2]$ and 
$[1/2,1]$. 
The corresponding Mellin convolutions with parton distribution functions of support $[0,1]$ are given by, 
cf.~\cite{Ablinger:2014nga},
\begin{eqnarray}
\left[A_1(x) \theta\left(\tfrac{1}{2} - x\right)\right] \otimes f(x) &=& \theta\left(\tfrac{1}{2} - x\right) 
\int_{2x}^1 \frac{dy}{y} A_1\left(\frac{x}{y}\right) f(y)
\\ 
\left[A_2(x) \theta\left(x - \tfrac{1}{2}\right)\right] \otimes f(x) &=& \int_x^1 \frac{dy}{y} A_2\left(\frac{x}{y}\right)
f(y) - \theta\left(\tfrac{1}{2} - x\right) 
\int_{2x}^1 \frac{dy}{y} A_2\left(\frac{x}{y}\right) f(y)~.
\end{eqnarray}

We will split $a_{Qq}^{(3),\rm PS}(x)$ into a part represented by the harmonic 
polylogarithms of only the argument $x$ and a part containing also harmonic polylogarithms with the argument $(1 - 2 x)$. 
In what follows, we use the shorthand notation
\begin{eqnarray}
H_{i_n, \cdots, i_1} = H_{i_n, \cdots, i_1}(x) 
\quad~~\text{and}~~\quad
\tilde{H}_{i_n, \cdots, i_1} = H_{i_n, \cdots, i_1}(1 - 2 x).
\end{eqnarray}
One obtains

\begin{eqnarray}
\label{eqaQq3}
\lefteqn{a_{Qq}^{(3), \rm PS}(x) =} \nonumber\\ &&
\textcolor{blue}{C_F T_F^2 N_F} \Biggl\{
        \frac{32}{3} (1-x) \biggl[
                \frac{2276}{81}
                -\biggl(
                        \frac{4}{3} H_1
                        +5 H_1^2
                \biggr) H_0
                +\biggl(
                        \frac{536}{27}
                        +\frac{25}{2} \zeta_2
                \biggr) H_1
                +10 H_0^2 H_1
                +\frac{1}{9} H_1^2
\nonumber\\ &&
                +\frac{5}{18} H_1^3
        \biggr]
        -\frac{128}{9} \biggl[
                \frac{2}{27} (104-229 x)
                +(10-11 x) H_{0,1}
        \biggr] H_0
        -\frac{32}{27} (5-4 x) \biggl(
                \frac{4}{3} H_0^3
                +7 \zeta_2 H_0
        \biggr)
\nonumber\\ &&
        +\frac{64}{9} (x+1) \biggl[
                \big(
                        15 \zeta_3
                        -12 H_{0,0,1}
                        -6 H_{0,1,1}
                \big) H_0
                +\biggl(
                        6 H_{0,1}
                        -\frac{7}{2} \zeta_2
                \biggr) H_0^2
                -\frac{1}{3} H_0^4
                +4 H_{0,0,0,1}
\nonumber\\ &&
                +10 H_{0,0,1,1}
                +H_{0,1,1,1}
                +\frac{15}{2} \zeta_2 H_{0,1}
                -\frac{21}{2} \zeta_2^2
        \biggr]
        -\frac{128}{81} (95 x+23) H_0^2
        +\frac{256}{81} (43 x+16) H_{0,1}
\nonumber\\ &&
        +\frac{640}{27} (4-5 x) H_{0,0,1}
        +\frac{64}{27} (40-41 x) H_{0,1,1}
        -\frac{32}{81} (389 x+83) \zeta_2
        +\frac{32}{27} (17 x+5) \zeta_3
\Biggr\}
\nonumber\\ &&
+\textcolor{blue}{C_F T_F^2} \Biggl\{
        \frac{64}{81} \big(
                12 x^4-15 x^3-x-46\big) H_0^2
        -\frac{128}{81} \big(
                24 x^4-30 x^3+214 x-65\big) H_{0,1}
\nonumber\\ &&
        +\frac{32}{3} (1-x) \biggl[
                \frac{8}{81} \big(
                        36 x^2+9 x-20\big)
                +\frac{1}{9} \big(
                        8 x^3-2 x^2-2 x+51\big) H_1^2
                -\frac{5}{9} H_1^3
\nonumber\\ &&
                -\biggl(
                        \frac{4}{9} \big(4 x^3-x^2-x+25\big) H_1
                        +5 H_1^2
                \biggr) H_0
                +\biggl(
                         20 H_{0,1}
                        -\frac{4}{27} \big(12 x^2+3 x+52\big)
                        -15 \zeta_2
                \biggr) H_1
        \biggl]
\nonumber\\ &&
        +\frac{32}{9} \biggl[
                -\frac{4}{27} \big(36 x^3-27 x^2-281 x+61\big)
                -4 (5 x+2) H_{0,1}
                -\frac{1}{3} (17-73 x) \zeta_2
        \biggr] H_0
\nonumber\\ &&
        +\frac{64}{3} (x+1) \biggl[
                2 \big(
                        -H_{0,1,1}
                        +3 \zeta_3
                \big) H_0
                -\frac{1}{9} H_0^4
                +2 H_{0,1}^2
                +\frac{4}{3} H_{0,0,0,1}
                -\frac{2}{3} H_{0,0,1,1}
\nonumber\\ &&
                -\frac{2}{3} H_{0,1,1,1}
                -\frac{7}{6} \zeta_2 H_0^2
                -3 \zeta_2 H_{0,1}
                +\frac{4}{5} \zeta_2^2
        \biggr]
        -\frac{128}{81} (5-4 x) H_0^3
        +\frac{128}{27} (26 x+17) H_{0,0,1}
\nonumber\\ &&
        -\frac{256}{27} (23-40 x) H_{0,1,1}
        +\frac{32}{81} \big(48 x^4-60 x^3+481 x+103\big) \zeta_2
        -\frac{64}{27} (317 x-163) \zeta_3
\Biggr\}
\nonumber\\ &&
+\textcolor{blue}{C_F^2 T_F} \Biggl\{
        -32 B_4 (5 x+13)
        -\frac{2}{x-1} \big(4 x^3-297 x^2+134 x+143\big) H_0^2
\nonumber\\ &&
        +\frac{32}{3}
         (1-x) \biggl\{
                \frac{29}{24} H_1^3
                +\frac{5}{48} H_1^4
                +\biggl[
                         \frac{3 x}{2} 
                        -\frac{319}{4} 
                        -70 H_{0,1}
                        +\frac{35}{2} \zeta_2                        
                        -\frac{21}{4} H_1
                        -\frac{5}{6} H_1^2
                \biggr] H_1 H_0
\nonumber\\ &&
                +\biggl(
                        \frac{471}{8} H_1
                        -\frac{5}{2} H_1^2
                \biggr) H_0^2
                +\biggl(
                         \frac{25}{2} \zeta_3
                        -\frac{405}{4}
                        +140 H_{0,0,1}
                        +95 H_{0,1,1}
                \biggr) H_1
                -\frac{25}{6} H_0^3 H_1
                -74
\nonumber\\ &&
                +\biggl[
                         \frac{451}{8}
                        -\frac{3 x}{4}
                        -15 H_{0,1}
                        +\frac{155}{8} \zeta_2
                \biggr] H_1^2
        \biggr\}
        +\frac{32}{3} (x+1) \biggl[
                -24 B_5
                -\biggl(
                        \frac{5}{3} H_{0,1}
                        +\frac{47}{24} \zeta_2
                \biggr) H_0^3
\nonumber\\ &&
                +\big(
                        -14 H_{0,1}^2
                        -20 H_{0,0,0,1}
                        +56 H_{0,0,1,1}
                        -2 H_{0,1,1,1}
                        +7 \zeta_2 H_{0,1}
                        -17 \zeta_2^2
                \big) H_0
                +\frac{9}{80} H_0^5
\nonumber\\ &&
                +\biggl(
                        \frac{9}{2} H_{0,0,1}
                        -2 H_{0,1,1}
                        -\frac{31}{4} \zeta_3
                \biggr) H_0^2
                +\big(
                        56 H_{0,0,1}
                        -12 H_{0,1,1}
                        +5 \zeta_3
                \big) H_{0,1}
                +80 H_{0,0,0,0,1}
\nonumber\\ &&
                -318 H_{0,0,0,1,1}
                -134 H_{0,0,1,0,1}
                +78 H_{0,0,1,1,1}
                +50 H_{0,1,0,1,1}
                +H_{0,1,1,1,1}
                +19 \zeta_2 H_{0,0,1}
\nonumber\\ &&
                +\frac{31}{2} \zeta_2 H_{0,1,1}
        \biggr]
        + \frac{2}{3} \big[
                4 (50 x+151)
                +8 (493 x-392) H_{0,1}
                -32 (67 x+127) H_{0,0,1}
\nonumber\\ &&
                -16 (131 x-134) H_{0,1,1}
                -(709 x-471) \zeta_2
                +8 (97 x-112) \zeta_3
        \big] H_0
        +\frac{2}{9} (253-15 x) H_0^3
\nonumber\\ &&
        +\biggl[
                \frac{16}{3} (61 x+104) H_{0,1}
                +\frac{20}{3} (7 x+13) \zeta_2
        \biggr] H_0^2
        +\frac{2}{9} (43-39 x) H_0^4
        +\frac{16}{3} (55 x-97) H_{0,1}^2
\nonumber\\ &&
        +\big[
                32 (23 x-21) H_{0,1}
                -4 (241 x-225) \zeta_2
        \big] H_1
        -\frac{8}{3} (1425 x-953) H_{0,0,1}
        -7168 \ln(2) \zeta_3
\nonumber\\ &&
        -\frac{32}{3} (255 x-209) H_{0,1,1}
        +\biggl[
                \frac{8}{3} \big(
                        12 x^2-1007 x+133\big)
                +120 (x+3) \zeta_2
        \biggr] H_{0,1}
\nonumber\\ &&
        +\frac{32}{3} (183 x+464) H_{0,0,0,1}
        -\frac{32}{3} (11-180 x) H_{0,0,1,1}
        -\frac{16}{3} (369-377 x) H_{0,1,1,1}
\nonumber\\ &&
        -\frac{2}{3} \big(24 x^2-3169 x-303\big) \zeta_2
        -\frac{8}{15} (897 x+2213) \zeta_2^2
        -\frac{4}{3} (1995-1291 x) \zeta_3
\Biggr\}
\nonumber\\ &&
+\textcolor{blue}{C_A C_F T_F} \Biggl\{
        16 B_4 (5 x+13)
        -\frac{8}{81 (x+1)} \big(
                162 x^3-29126 x^2-4171 x+24793\big) H_{0,1}
\nonumber\\ &&
        -\frac{4}{81 (1-x^2)}
        \big(108 x^4+58694 x^3+2948 x^2-59261 x-3299\big) H_0^2
\nonumber\\ &&
        +\frac{32}{3} (43 x+41) \big(
                H_{0,-1,1}
                +H_{0,1,-1}
        \big) H_0
        -\frac{8}{3} \biggl[
                \frac{1}{9} (14221-2825 x)
                +2 (3 x+7) \zeta_2
        \biggr] H_{0,0,1}
\nonumber\\ &&
        -16 (63-79 x) \biggl[
                H_{0,0,0,-1,1}
                +H_{0,0,0,1,-1}
                +\frac{1}{3} H_{0,0,1,0,-1}
        \biggr]
        -\frac{16}{3} (197 x+411) H_{0,0,0,0,1}
\nonumber\\ &&
        +\frac{32}{3} (1-x) \biggl\{
                \biggl(
                         \frac{19}{4} H_{0,-1,-1}
                        -\frac{185}{16} H_1^2
                \biggr) H_0^2
                -\biggl[
                        \frac{1}{72} (1123-27 x)
                        +\frac{25}{8} \zeta_2
                \biggr] H_1^2
                -\frac{128383}{81}
\nonumber\\ &&
                +\biggl[
                        \biggl(
                                \frac{1}{12} (2996-9 x)
                                +H_{0,-1}
                                +\frac{239}{4} H_{0,1}
                                -3 \zeta_2
                        \biggr) H_1
                        +29 H_1^2
                        +\frac{10}{3} H_1^3
                        +\frac{19}{4} H_{0,-1}^2
\nonumber\\ &&
                        +14 H_{0,-1,-1,-1}
                        -30 H_{0,0,-1,-1}
                        -\frac{73}{2} H_{0,0,0,-1}
                        +\frac{19}{2} \zeta_2 H_{0,-1}
                \biggr] H_0
                -\frac{115}{72} H_1^3
                -17 \zeta_2 H_{0,0,-1}
\nonumber\\ &&
                +\biggl(
                        \frac{169}{24} H_1
                        -\frac{35}{12} H_{0,-1}
                \biggr) H_0^3
                +\biggl(
                        \frac{77}{108}
                        -2 H_{0,0,-1}
                        -\frac{361}{4} H_{0,0,1}
                        -\frac{55}{2} H_{0,1,1}
                        -\frac{73}{4} \zeta_3
                \biggr) H_1
\nonumber\\ &&
                +\biggl(
                        H_{0,1}
                        -14 H_{0,-1,-1}
                        -\frac{57}{2} H_{0,0,-1}
                        +\frac{71}{2} H_{0,0,1}
                        -19 \zeta_3
                \biggr) H_{0,-1}
                -5 H_{0,-1,-1,0,1}
                -\frac{5}{48} H_1^4
\nonumber\\ &&
                +28 H_{0,-1,0,-1,-1}
                +56 H_{0,0,-1,-1,-1}
                +\frac{117}{2} H_{0,0,-1,0,-1}
                +\frac{351}{2} H_{0,0,0,-1,-1}
                +12 \zeta_2 H_{0,-1,-1}
        \biggr\}
\nonumber\\ &&
        +\frac{32}{3} (x+1) \biggl\{
                12 B_5
                +\frac{1}{4 x} \big(x^2-908 x+3\big) \big(H_{0,-1}-H_{-1} H_0\big)
                +\frac{35}{6} H_{-1}^3 H_0
                +\frac{31}{12} H_0^3 H_{0,1}
\nonumber\\ &&
                -\biggl(
                        4 H_{0,0,-1}
                        +\frac{71}{2} H_{0,0,1}
                        +\frac{11}{2} \zeta_3
                \biggr) H_{0,1}
                +\biggl[
                        \biggl(
                                \frac{77}{4} H_{0,-1}
                                -42 H_{0,1}
                                +\frac{131}{4} \zeta_2
                        \biggr) H_0
                        -\frac{63}{8} H_0^3
\nonumber\\ &&
                        +62 H_{0,-1}
                        +35 H_{0,-1,-1}
                        -\frac{1}{2} H_{0,-1,1}
                        -\frac{271}{4} H_{0,0,-1}
                        +\frac{345}{4} H_{0,0,1}
                        -\frac{1}{2} H_{0,1,-1}
                        -2 H_{0,1,1}
\nonumber\\ &&
                        -\frac{75}{2} \zeta_3
                \biggr] H_{-1}
                +\biggl(
                        -31 H_0
                        +\frac{117}{16} H_0^2
                        -\frac{35}{2} H_{0,-1}
                        +\frac{1}{4} H_{0,1}
                        +\frac{17}{2} \zeta_2
                \biggr) H_{-1}^2
                +\biggl(
                        2 H_{0,-1} H_{0,1}
\nonumber\\ &&
                        +\frac{49}{4} H_{0,1}^2
                        -4 H_{0,0,-1,1}
                        -4 H_{0,0,1,-1}
                        +8 H_{0,1,1,1}
                        -\zeta_2 H_{0,1}
                \biggr) H_0
                -62 H_{0,-1,-1}
                -\frac{39}{4} H_0^2 H_{0,1,1}
\nonumber\\ &&
                -35 H_{0,-1,-1,-1}
                +\frac{1}{2} H_{0,-1,-1,1}
                +\frac{1}{2} H_{0,-1,1,-1}
                +2 H_{0,-1,1,1}
                +\frac{271}{4} H_{0,0,-1,-1}
                -\frac{345}{4} H_{0,0,-1,1}
\nonumber\\ &&
                -\frac{345}{4} H_{0,0,1,-1}
                +\frac{1}{2} H_{0,1,-1,-1}
                +2 H_{0,1,-1,1}
                +2 H_{0,1,1,-1}
                -15 H_{0,0,1,1,1}
                -11 H_{0,1,0,1,1}
\nonumber\\ &&
                -H_{0,1,1,1,1}
                -\frac{5}{2} \zeta_2 H_{0,1,1}
        \biggr\}
        +\frac{4}{3} \biggl[
                (173 x+135) H_{0,-1}
                -(247 x+887) H_{0,1}
                +\frac{\zeta_2}{3} (241 x+223)
\nonumber\\ &&
                -3 (436-431 x) H_1
                +8 (15-17 x) H_{0,0,-1}
                -4 (25 x+69) H_{0,0,1}
                +4 (3 x-41) \zeta_3
        \biggr] H_0^2
\nonumber\\ &&
        +\frac{8}{3} \biggl[
                23 (15-14 x) H_{0,-1}
                +\frac{1}{3} (4322-2473 x) H_{0,1}
                -(207 x+319) H_{0,-1,-1}
\nonumber\\ &&
                -(59 x+87) H_{0,0,-1}
                +7 (59 x+209) H_{0,0,1}
                +120 (3 x-2) H_{0,1,1}
                -8 (9-7 x) H_{0,-1,0,1}
\nonumber\\ &&
                +186 (x+3) H_{0,0,0,1}
                -2 (37 x+35) H_{0,0,1,1}
                +\frac{1}{18} (3629-2707 x) \zeta_2
                +\frac{1}{5} (377-231 x) \zeta_2^2
\nonumber\\ &&
                -2 (179 x+66) \zeta_3
                -\frac{1}{81} (433786 x+86449)
        \biggr] H_0
        +\frac{4}{15} (4-5 x) H_0^5
        +32 (x+73) \zeta_5
\nonumber\\ &&
        +\biggl[
                -4 (42 x+41) H_0^2
                +\frac{16}{3} (10 x+7) H_{0,1}
                -16 (24 x+23) \zeta_2
        \biggr] H_{-1}
        +\frac{2}{27} (292 x+223) H_0^4
\nonumber\\ &&
        +\biggl[
                \frac{4}{81} (1733-1669 x)
                -\frac{16}{9} (4 x-5) \zeta_2
        \biggr] H_0^3
        +4 \biggl[
                \frac{4}{3} (53-56 x) H_{0,1}
                -(181-185 x) \zeta_2
        \biggr] H_1
\nonumber\\ &&
        +\frac{8}{3} (65 x+121) H_{0,-1}^2
        +\frac{4}{3} (209-83 x) H_{0,1}^2
        -\frac{16}{3} (10 x+7) \big(H_{0,-1,1}+H_{0,1,-1}\big)
\nonumber\\ &&
        -\frac{56}{3} (81-110 x) H_{0,0,-1}
        +\frac{8}{27} (6155 x-4393) H_{0,1,1}
        -\frac{8}{3} (157 x+169) H_{0,-1,0,1}
\nonumber\\ &&
        +24 (5-17 x) H_{0,0,0,-1}
        -\frac{16}{9} (491 x+2906) H_{0,0,0,1}
        -\frac{16}{9} (851 x+47) H_{0,0,1,1}
\nonumber\\ &&
        +\frac{16}{9} (358-395 x) H_{0,1,1,1}
        -\frac{32}{3} (33-41 x) H_{0,0,-1,0,1}
        -128 (x-3) H_{0,0,0,0,-1}
\nonumber\\ &&
        +\frac{16}{3} (347 x+341) H_{0,0,0,1,1}
        +\frac{32}{3} (67 x+68) H_{0,0,1,0,1}
        -\frac{8}{3} (115 x+167) \zeta_2 H_{0,-1}
\nonumber\\ &&
        -\frac{8}{3} (106 x+121) \zeta_2 H_{0,1}
        +\frac{4}{15} (3163 x+7429) \zeta_2^2
        +\biggl[3584 \ln(2)+\frac{136}{27} (100 x+697)\biggr] \zeta_3
\nonumber\\ &&
        -\frac{16}{3} (45-34 x) \zeta_2 \zeta_3
        +\frac{4 \zeta_2}{81 (x+1)} \big(108 x^3+9131 x^2+40690 x+31019\big)
\Biggr\}
+ \tilde{a}_{Qq}^{(3),\rm PS},
\end{eqnarray}

where
\begin{eqnarray}
\label{eqatildeQq3}
\lefteqn{\tilde{a}_{Qq}^{(3), \rm PS}(x) =} \nonumber\\ &&
\textcolor{blue}{C_F T_F \left(C_A-2 C_F\right)}
\Biggl\{
        -32 (11-23 x) \big[
                2 \ln(2) \big(
                        \tilde{H}_{0,-1}
                        +\tilde{H}_{0,1}
                \big)
                -\tilde{H}_{0,-1,-1}
                +\tilde{H}_{0,-1,1}
\nonumber\\ &&
                -\tilde{H}_{0,1,-1}
                +\tilde{H}_{0,1,1}
        \big]
        +160 (1-x) \big[
                -4 \ln(2) \tilde{H}_{0,-1,-1}
                +3 \tilde{H}_{0,-1,-1,-1}
                -\tilde{H}_{0,1,-1,1}
        \big]
\nonumber\\ &&
        +32 (5 x+7) \big[
                -4 \ln(2) \tilde{H}_{0,1,1}
                +\tilde{H}_{0,-1,1,-1}
                -3 \tilde{H}_{0,1,1,1}
        \big]
        +96 (5 x+3) \tilde{H}_{0,1,1,-1}
\nonumber\\ &&
        +64 (x+1) \biggl\{
                \ln(2) \biggl[
                        \big(
                                -2 \tilde{H}_{-1} \tilde{H}_1
                                -\tilde{H}_1^2
                                +2 \tilde{H}_{-1,1}
                        \big) \big(\tilde{H}_{0,-1}+\tilde{H}_{0,1}\big)
                        -4 \tilde{H}_{0,-1,-1,1}
\nonumber\\ &&
                        +4 \biggl(
                                \tilde{H}_{0,-1,-1}
                                +\tilde{H}_{0,-1,1}
                                +\tilde{H}_{0,1,-1}
                                +\tilde{H}_{0,1,1}
                                -\frac{7}{2} \zeta_3
                        \biggr) \tilde{H}_1
                        -2 \tilde{H}_{0,-1,1,-1}
                        -6 \tilde{H}_{0,-1,1,1}
\nonumber\\ &&
                        -4 \tilde{H}_{0,1,-1,1}
                        -2 \tilde{H}_{0,1,1,-1}
                        -6 \tilde{H}_{0,1,1,1}
                \biggr]
                +2 \tilde{H}_{0,1,-1,1,-1}
                -3 \tilde{H}_{0,1,-1,1,1}
                +\tilde{H}_{0,1,1,-1,-1}
\nonumber\\ &&
                +\biggl(
                        \tilde{H}_{-1} \tilde{H}_1+
                        \frac{1}{2} \tilde{H}_1^2
                        -\tilde{H}_{-1,1}
                \biggr)
\biggl(\tilde{H}_{0,-1,-1}
                        -\tilde{H}_{0,-1,1}
                        +\tilde{H}_{0,1,-1}
                        -\tilde{H}_{0,1,1}
                        +\frac{7}{2} \zeta_3
                \biggr)
\nonumber\\ &&
                +\biggl(
                        -B_4
                        -3 \tilde{H}_{0,-1,-1,-1}
                        +\tilde{H}_{0,-1,-1,1}
                        -\tilde{H}_{0,-1,1,-1}
                        +3 \tilde{H}_{0,-1,1,1}
                        -3 \tilde{H}_{0,1,-1,-1}
\nonumber\\ &&
                        +\tilde{H}_{0,1,-1,1}
                        -\tilde{H}_{0,1,1,-1}
                        +3 \tilde{H}_{0,1,1,1}
                        +\frac{9}{20} \zeta_2^2
                \biggr) \tilde{H}_1
                +3 \tilde{H}_{0,-1,-1,-1,1}
                +2 \tilde{H}_{0,-1,-1,1,-1}
\nonumber\\ &&
                -3 \tilde{H}_{0,-1,-1,1,1}
                +\tilde{H}_{0,-1,1,-1,-1}
                +3 \tilde{H}_{0,-1,1,1,-1}
                -6 \tilde{H}_{0,-1,1,1,1}
                +3 \tilde{H}_{0,1,-1,-1,1}
\nonumber\\ &&
                +3 \tilde{H}_{0,1,1,1,-1}
                -6 \tilde{H}_{0,1,1,1,1}
        \biggr\}
        -32 (5 x+19) \tilde{H}_{0,-1,1,1}
        -32 (5 x-17) \tilde{H}_{0,1,-1,-1}
\nonumber\\ &&
        +32 \big(
                5 (1-x) \tilde{H}_{-1}
                +(5 x+7) \tilde{H}_1
        \big)
\biggl[2 \ln(2) \big(
                        \tilde{H}_{0,-1}
                        +\tilde{H}_{0,1}
                \big)
                -\tilde{H}_{0,-1,-1}
                +\tilde{H}_{0,-1,1}
\nonumber\\ &&
                -\tilde{H}_{0,1,-1}
                +\tilde{H}_{0,1,1}
                -\frac{7}{2} \zeta_3
        \biggr]
        -768 \ln(2) \big(
                \tilde{H}_{0,-1,1}
                +\tilde{H}_{0,1,-1}
        \big)
        +96 (5 x-1) \tilde{H}_{0,-1,-1,1}
\Biggr\}.
\end{eqnarray}

Let us now derive expansions in the small and large $x$ regions for $a_{Qq}^{\rm PS, (3)}(x)$.
For small values of $x$ the following approximation holds
\begin{eqnarray}
\label{eq:SX1}
a_{Qq}^{\rm PS, (3)}(x) &\simeq&
\textcolor{blue}{C_F T_F^2}
\Biggl\{
        -\frac{64}{27} \ln^4(x)
        -\frac{640}{81} \ln^3(x)
        -\frac{32}{81} \left(
                92
                +63 \zeta_2
        \right) \ln^2(x)
\nonumber\\ &&        
-
        \frac{32}{243} \left(
                244
                +153 \zeta_2
                -972 \zeta_3
        \right) \ln(x)
        +\frac{32}{1215} \left(
                -800
                +1545 \zeta_2
                +648 \zeta_2^2
                +14670 \zeta_3
        \right)
\nonumber\\ &&
        +\textcolor{blue}{N_F} \Biggl[
                -\frac{64}{27}  \ln^4(x)
                -\frac{640}{81} \ln^3(x)
                -\frac{32}{81} \left(
                        92
                        +63 \zeta_2
                \right) \ln^2(x)
\nonumber\\ &&
               -\frac{32}{243} \left(
                        832
                        +315 \zeta_2
                        -810 \zeta_3
                \right) \ln(x)
                -\frac{32}{243} \left(
                        -2276
                        +249 \zeta_2
                        +567 \zeta_2^2
                        -45 \zeta_3
                \right)
        \Biggr] \Biggr\}
\nonumber\\ &&
+ \textcolor{blue}{C_F C_A T_F} \Biggl\{
         \frac{16}{15} \ln^5(x)
        +\frac{446}{27} \ln^4(x)
        +\frac{4}{81} \big(
                1733
                +180 \zeta_2
        \big) \ln^3(x)
\nonumber\\ &&
        +\frac{4}{81} \big(
                3299
                +2007 \zeta_2
                -4428 \zeta_3
        \big) \ln^2(x)
        +\frac{4}{1215} \big(
                -866920
                +163305 \zeta_2
\nonumber\\ &&
                +61074 \zeta_2^2
                -106920 \zeta_3
        \big) \ln(x)
+ 32 B_4
        +\frac{4}{1215} \big(
                -5132890
                +465285 \zeta_2
\nonumber\\ &&               
 +593001 \zeta_2^2
                +692190 \zeta_3
                -93150 \zeta_2 \zeta_3
                +660960 \zeta_5
        \big)
\Biggr\}
\nonumber\\ &&
+\textcolor{blue}{C_F^2 T_F} \Biggl\{
\frac{6}{5} \ln^5(x) 
+ \frac{86}{9} \ln^4(x)  
     + \frac{2}{9} \big(
                253
                -94 \zeta_2
        \big) \ln^3(x)
\nonumber\\ &&
        +\frac{2}{3} \big(
                429
                +130 \zeta_2
                -124 \zeta_3
        \big) \ln^2(x)
        +\frac{2}{3} \big(
                604
                +471 \zeta_2
                -272 \zeta_2^2
                -896 \zeta_3
        \big) \ln(x)
\nonumber\\ &&              
        -64 B_4
        -\frac{2}{3} \big(
                1184
                -303 \zeta_2
                +1684 \zeta_2^2
  +294 \zeta_3
                -200 \zeta_2 \zeta_3
                -480 \zeta_5
        \big)
\Biggr\}.
\end{eqnarray}
Here the leading terms is
\begin{eqnarray}
\label{eq:SX0}
a_{Qq}^{\rm PS, (3)}(x) &\simeq& \frac{2}{15} \textcolor{blue}{ C_F T_F} \left[
                                 8 \textcolor{blue}{C_A} + 9 \textcolor{blue}{C_F} \right]
         \ln^5(x).
\end{eqnarray}

For large values of $x$ one obtains
\begin{eqnarray}
\label{eq:LX}
a_{Qq}^{\rm PS, (3)}(x) &\simeq&
(1-x) \Biggl\{
\textcolor{blue}{C_F T_F^2 N_F} \Biggl[
        -\frac{16}{27} \ln^3(1-x)
        -\frac{128}{27} \ln^2(1-x)
        -\frac{16}{81} \big(
                68
                +135 \zeta_2
        \big) \ln(1-x)
\nonumber\\ &&    
    -\frac{32}{243} \big(
                230
                +540 \zeta_2
                -297 \zeta_3
        \big)
\Biggr]
\nonumber\\ &&
+\textcolor{blue}{C_F T_F^2} \Biggl[
         \frac{32}{27} \ln^3(1-x)
        -\frac{32}{27} \ln^2(1-x)
        -\frac{32}{81} \big(
                -32
                +27 \zeta_2
        \big) \ln(1-x)
\nonumber\\ &&
        -\frac{32}{243} \big(
                -64
                +459 \zeta_2
                -378 \zeta_3
        \big)
\Biggr]
\nonumber\\ &&
+ \textcolor{blue}{C_A C_F T_F} \Biggl[
        -\frac{2}{9} \ln^4(1-x)
        +\frac{68}{27} \ln^3(1-x)
        -\frac{4}{27} \big(
                -136
                +45 \zeta_2
        \big) \ln^2(1-x)
\nonumber\\ &&
+        \frac{4}{81} \big(
                1120
                +2403 \zeta_2
                +1404 \zeta_3
        \big) \ln(1-x)
        +\frac{4}{1215} \big(
                49135
                +84915 \zeta_2
\nonumber\\ &&
                +7128 \zeta_2^2
                -77220 \zeta_3
        \big)
\Biggr]
+\textcolor{blue}{C_F^2 T_F} \Biggl[
         \frac{2}{9} \ln^4(1-x)
        -\frac{20}{9} \ln^3(1-x)
\nonumber\\ &&
        +\frac{4}{3} \big(
                -19
                +7 \zeta_2
        \big) \ln^2(1-x)
        -\frac{4}{3} \big(
                34
                +91 \zeta_2
                +60 \zeta_3
        \big) \ln(1-x)
\nonumber\\ &&
        -\frac{2}{3} \big(
                252
                +335 \zeta_2
                +168 \zeta_2^2
                -278 \zeta_3
        \big)
\Biggr] \Biggr\}~.
\end{eqnarray}
\begin{figure}[ht]
  \centering
  \hskip-0.8cm
  \includegraphics[width=.7\linewidth]{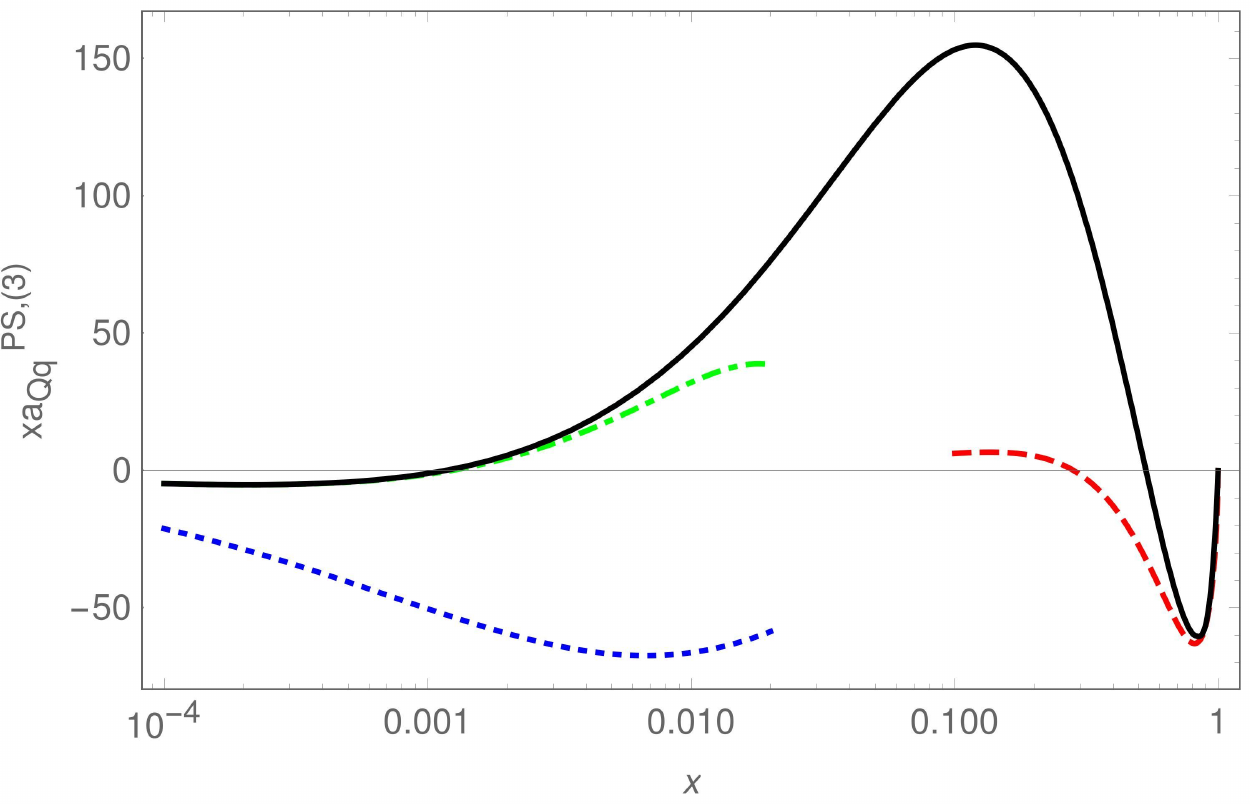}
  \caption[]{\sf The constant part of the polarized unrenormalized OME $A_{Qq}^{\rm PS,(3)}$, $x a_{Qq}^{\rm PS,(3)}$, as a function 
of 
$x$. Full line: complete expression; dotted line: leading small $x$ term (\ref{eq:SX0}); dash-dotted line: small $x$ approximation 
(\ref{eq:SX1}); dashed line: large $x$ approximation (\ref{eq:LX}) for $N_F$=3.}
  \label{fig:plot1}
\end{figure}

\noindent
In Figure~\ref{fig:plot1} we illustrate the function $x a_{Qq}^{\rm PS,(3)}$ setting $N_F = 3$. The leading small $x$ term is 
phenenomenologically not dominant, as has also been observed for a large number of other quantities, cf.~\cite{Blumlein:1997em,Blumlein:1996hb,Blumlein:1999ev}. 
A series of 
sub-leading terms is necessary to describe $x a_{Qq}^{\rm PS,(3)}$ at least for some region in the small $x$ domain. The same is true
in the large $x$ region. In both cases the complete logarithmic approximations either in $\ln^k(x)$ or $\ln^k(1-x)$ down to the 
constant term have a limited range of validity only and the complete function mainly depends on other structures.

\vspace*{3mm}
Let us now turn to the complete OME $A_{Qq}^{(3), \rm PS}$.
From Eq. (\ref{AQq3PSMSren}), we obtain the following renormalized result in $N$ space,

\begin{eqnarray}
A_{Qq}^{(3), \rm PS} &=& 
\textcolor{blue}{C_F^2 T_F} \Biggl\{
         \frac{4}{3} F_1 \big(
                 F_3
                 -4 S_1
         \big) L^3
         +8 L^2 \biggl[
                 -\frac{(N-1)^2 (N+2) (3 N+2)}{N^3 (N+1)^3} S_1
\nonumber \\ &&
                 +F_1 \biggl(
                         \frac{P_{40}}{N^2 (N+1)^2}
                         +2 S_2
                 \biggr)
         \biggr]
         +L \biggl[
                 -\frac{4 P_{61}}{N^5 (N+1)^5}
                 +32 F_1 \biggl(
                         -\frac{1}{12} S_1^3
                         -\frac{1}{6} S_3
\nonumber \\ &&
                         -\frac{1}{4} S_1 S_2
                         -S_{1,2}
                         +S_{1,1,1}
                         +3 \zeta_3
                 \biggr)
                 +\frac{8 P_{54} S_1}{N^4 (N+1)^4}
                 +\frac{8 P_{46} S_{1,1}-8 P_{42} S_1^2}{N^3 (N+1)^3}
\nonumber \\ &&
                 -\frac{8 (N-1) P_{37} S_2}{N^3 (N+1)^3}
         \biggr]
         -\frac{4 P_{64}}{N^6 (N+1)^6}
         +F_1 \biggl[
                 32 \biggl(
                         \frac{1}{6} S_3
                         +\frac{3 N^2-3 N-4}{8 N (N+1)} \zeta_2
\nonumber \\ &&
                         +S_{2,1}
                         +\frac{\zeta_3}{6}
                 \biggr) S_1 
                 +4 \big(
                         S_2
                         +2 \zeta_2
                 \big) S_1^2
                 +\frac{2}{3} S_1^4
                 +2 S_2^2
                 -12 S_4
                 +32 S_{3,1}
                 -8 \zeta_2 S_2
\nonumber \\ &&
                 -8 \zeta_2 S_{1,1}
                 -64 S_{2,1,1}
                 -\frac{2 P_{47} \zeta_2}{N^2 (N+1)^2}
                 -\frac{4}{3} \zeta_3 F_3
         \biggr]
         -F_4 \biggl(
                 \frac{8}{3} S_1^3
                 +8 S_1 S_2
         \biggr)
\nonumber \\ &&
         +\frac{24 P_{39} S_1-8 P_{41} S_3}{3 N^3 (N+1)^3}
         -\frac{4 P_{31} S_1^2}{N^3 (N+1)^2}
         -\frac{4 P_{55} S_2}{N^4 (N+1)^4}
         +\frac{32 (N+2) S_{2,1}}{N^3 (N+1)}
 \Biggr\}
\nonumber \\ &&
 +\textcolor{blue}{C_F T_F^2} \Biggl\{
         -\frac{128}{9} F_1 L^3
         +\frac{32}{3} L^2 \biggl(
                 -\frac{F_2}{3}
                 +F_1 S_1
         \biggr)
         +\frac{32}{3} L \biggl[
                 -\frac{2 (N+2) P_{52}}{9 N^4 (N+1)^4}
\nonumber \\ &&
                 -F_1 \biggl(
                         \frac{1}{2} S_1^2
                         +\frac{5}{2} S_2
                         +S_{1,1}
                 \biggr)
                 +\frac{2 (N+2) P_{34} S_1}{3 N^3 (N+1)^3}
         \biggr]
         +\frac{32 (N+2) P_{60}}{81 N^5 (N+1)^5}
         +\frac{32}{9} \zeta_2 F_2
\nonumber \\ &&
         +\frac{32}{9} F_1 \biggl[
                 \biggl(
                         -\frac{28 N^2+41 N+22}{3 (N+1)^2}
                         -
                         \frac{3}{2} S_2
                         -3 \zeta_2
                 \biggr) S_1
                 +\frac{(5 N+2) S_1^2}{2 (N+1)}
                 -\frac{1}{2} S_1^3
\nonumber \\ &&
                 +5 S_3
                 +4 \zeta_3
         \biggr]
         -\frac{16 (N+2) P_{33} S_2}{9 N^3 (N+1)^3}
 \Biggr\}
+\textcolor{blue}{C_F T_F^2 N_F} \Biggl\{
                 -\frac{32}{9} F_1 L^3
                 +\frac{32}{3} L^2 \biggl(
                         -F_1 S_1
\nonumber \\ &&
                         +\frac{(N+2) P_{34}}{3 N^3 (N+1)^3}
                 \biggr)
                 +\frac{32}{3} L \biggl[
                         -\frac{(N+2) P_{53}}{9 N^4 (N+1)^4}
                         +F_1 \biggl(
                                 \frac{1}{2} S_1^2
                                 -\frac{3}{2} S_2
                                 -2 S_{1,1}
                         \biggr)
\nonumber \\ &&
                         +\frac{(N+2) P_{36} S_1}{3 N^3 (N+1)^3}
                 \biggr]
                 -\frac{32 (N+2) P_{50}}{3 N^5 (N+1)^5}
                 +\frac{32}{3} F_1 \biggl(
                         2 S_3
                         +\frac{1}{2} \zeta_2 S_1
                         +\frac{\zeta_3}{3}
                 \biggr)
\nonumber \\ &&
                 -\frac{64 (N+2) \big(N^3+2 N+1\big)}{3 N^3 (N+1)^3} S_2
                 -\frac{16 (N+2) \zeta_2 P_{36}}{9 N^3 (N+1)^3}
         \Biggr\}
\nonumber \\ &&
 +\textcolor{blue}{C_A C_F T_F} \Biggl\{
         \frac{16}{3} F_1 \biggl(
                 \frac{11 N^2+11 N-12}{6 N (N+1)}
                 +S_1
         \biggr) L^3
         +8 L^2 \biggl[
                 -\frac{P_{57}}{9 N^4 (N+1)^4}
\nonumber \\ &&
                 +2 F_1 \big(
                         S_2
                         +2 S_{-2}
                 \big)
                 +\frac{P_{45} S_1}{3 N^3 (N+1)^3}
         \biggr]
         +8 L \biggl[
                 \frac{P_{62}}{27 N^5 (N+1)^5}
                 +4 F_1 \biggl(
                         S_{-2} S_1
\nonumber \\ &&
                         +\frac{1}{12} S_1^3
                         +\frac{3}{4} S_1 S_2
                         +S_{1,2}
                         +S_{1,-2}
                         -S_{1,1,1}
                         -3 \zeta_3
                 \biggr)
                 -\frac{P_{59} S_1}{9 N^4 (N+1)^4}
\nonumber \\ &&
                 +\frac{P_{35} S_1^2}{N^2 (N+1)^3}
                 +\frac{23 N^2+23 N-58}{3 N^2 (N+1)^2} S_3
                 +\frac{4 \big(3 N^2+3 N-4\big)}{N^2 (N+1)^2} S_{-3}
\nonumber \\ &&
                 -\frac{2 (N-2) P_{32} S_{-2}}{N^3 (N+1)^3}
                 +\frac{P_{49} S_2}{3 N^3 (N+1)^3}
                 -\frac{P_{48} S_{1,1}}{3 N^3 (N+1)^3}
                 -4 N F_4 S_{-2,1}
         \biggr]
\nonumber \\ &&
         +32 F_1 \biggl[
                 -S_1^2 \biggl(
                         \frac{5}{8} S_2
                         +\frac{\zeta_2}{4}
                 \biggr)
                 -S_{-2} \biggl(
                         \frac{2 S_1}{N+1}
                         +S_1^2
                         +S_2
                         +\frac{3 \zeta_2}{4}
                 \biggr)
                 -\frac{1}{48} S_1^4
\nonumber \\ &&
                 -\frac{1}{16} S_2^2
                 +\biggl(
                         -\frac{5}{3} S_3
                         +2 S_{-2,1}
                         -\frac{\zeta_3}{6}
                 \biggr) S_1
                 -\frac{9}{8} S_4
                 -\biggl(
                         \frac{1}{N+1}
                         +S_1
                 \biggr) S_{-3}
\nonumber \\ &&
                 -\frac{1}{2} S_{-4}
                 +\frac{1}{2} S_{3,1}
                 +\frac{2 S_{-2,1}}{N+1}
                 +S_{-2,2}
                 +S_{-3,1}
                 +\frac{1}{2} S_{2,1,1}
                 -2 S_{-2,1,1}
                 -\frac{1}{2} \zeta_2 S_2
\nonumber \\ &&
                 +\frac{1}{4} \zeta_2 S_{1,1}
                 -\frac{\big(11 N^2+11 N-12\big) \zeta_3}{36 N (N+1)}
         \biggr]
         +\frac{16 P_{63}}{3 N^6 (N+1)^6}
         -\biggl(
                  \frac{8 P_{51}}{N^2 (N+1)^5}
\nonumber \\ &&
                 +\frac{8 \big(3 N^2-13\big) S_2}{N^2 (N+1)^3}
                 +\frac{4 P_{44} \zeta_2}{3 N^3 (N+1)^3}
         \biggr) S_1
         +\frac{4 P_{38} S_1^2}{N^2 (N+1)^4}
         +\frac{4 P_{56} S_2}{3 N^4 (N+1)^4}
\nonumber \\ &&
         +\frac{8 \big(N^2+4 N+5\big)}{3 N^2 (N+1)^3} S_1^3
         -\frac{16 P_{43} S_3}{3 N^3 (N+1)^3}
         +\frac{32 (N+2) \big(N^2+3\big) S_{-2}}{N^2 (N+1)^4}
\nonumber \\ &&
         +\frac{4 P_{58} \zeta_2}{9 N^4 (N+1)^4}
 \Biggl\}
+a_{Qq}^{(3), \rm PS}(N).
\end{eqnarray}

Here $L=\ln\left(m^2/\mu^2\right)$. $F_1$ and $F_2$ were given in Eqs.~(\ref{F1inN}) and (\ref{F2inN}), and

\begin{eqnarray}
F_3 &=& \frac{3 N^2+3 N+2}{N (N+1)}, \\
F_4 &=& \frac{3 N^2+3 N-2}{N^3 (N+1)^2},
\end{eqnarray}

\noindent
The polynomials $P_i$ are
\begin{eqnarray}
  P_{31} &=& N^3-6 N^2-22 N-36, \\
  P_{32} &=& N^3-2 N^2-10 N-1, \\ 
  P_{33} &=& 7 N^3+3 N^2+26 N+12, \\
  P_{34} &=& 8 N^3-3 N^2+4 N+3, \\ 
  P_{35} &=& 10 N^3+13 N^2-4 N-5, \\
  P_{36} &=& 11 N^3-3 N^2+10 N+6, \\ 
  P_{37} &=& 13 N^3+18 N^2-13 N-14, \\
  P_{38} &=& N^4+2 N^3-5 N^2-12 N+2, \\ 
  P_{39} &=& 2 N^4-4 N^3-3 N^2+20 N+12, \\
  P_{40} &=& 2 N^4+N^3-3 N-2, \\ 
  P_{41} &=& 3 N^4+48 N^3+123 N^2+98 N+8, \\ 
  P_{42} &=& 10 N^4+11 N^3-6 N^2-N+2, \\ 
  P_{43} &=& 11 N^4+18 N^3-39 N^2-54 N+24, \\ 
  P_{44} &=& 11 N^4+22 N^3-23 N^2-70 N-12, \\ 
  P_{45} &=& 11 N^4+22 N^3-11 N^2-46 N-12, \\ 
  P_{46} &=& 17 N^4+22 N^3+N^2+4 N+4, \\ 
  P_{47} &=& 35 N^4+64 N^3+28 N^2-13 N-6, \\ 
  P_{48} &=& 49 N^4+62 N^3-13 N^2-2 N+12, \\ 
  P_{49} &=& 52 N^4+71 N^3-100 N^2-125 N+48, \\ 
  P_{50} &=& N^5-7 N^4+6 N^3+7 N^2+4 N+1, \\ 
  P_{51} &=& 2 N^5+10 N^4+29 N^3+64 N^2+67 N+8, \\ 
  P_{52} &=& 43 N^5+19 N^4+38 N^3-22 N^2-21 N-9, \\ 
  P_{53} &=& 58 N^5+25 N^4+95 N^3-22 N^2-42 N-18, \\ 
  P_{54} &=& 22 N^6+51 N^5+6 N^4-69 N^3-142 N^2-92 N-24, \\ 
  P_{55} &=& 27 N^6+102 N^5+131 N^4+52 N^3+20 N+8, \\ 
  P_{56} &=& 47 N^6+186 N^5+365 N^4+560 N^3+398 N^2-104 N-48, \\ 
  P_{57} &=& 118 N^6+321 N^5+97 N^4-27 N^3+253 N^2-114 N-72, \\ 
  P_{58} &=& 160 N^6+447 N^5+211 N^4+159 N^3+475 N^2-192 N-108, \\ 
  P_{59} &=& 169 N^6+474 N^5+355 N^4+240 N^3+547 N^2+321 N+18, \\ 
  P_{60} &=& 164 N^7+244 N^6-66 N^5-82 N^4-260 N^3-189 N^2-108 N-27, \\ 
  P_{61} &=& 36 N^8+110 N^7+98 N^6-4 N^5-149 N^4-188 N^3-187 N^2-128 N-36, \\
  P_{62} &=& 968 N^8+3473 N^7+4952 N^6+6113 N^5+3887 N^4-2512 N^3+705 N^2
\nonumber \\ && +1062 N+432, \\ 
  P_{63} &=& 6 N^{10}+33 N^9+73 N^8+32 N^7-88 N^6+38 N^5+241 N^4+87 N^3
\nonumber \\ && +29 N^2-13 N-6, \\ 
  P_{64} &=& 24 N^{10}+104 N^9+213 N^8+272 N^7+101 N^6-207 N^5-259 N^4
\nonumber \\ && -107 N^3-13 N^2+12 N+4.
\end{eqnarray}

The corresponding result in $x$ space is given by

\begin{eqnarray}
A_{Qq}^{(3), \rm PS} &=&
\textcolor{blue}{C_F T_F^2} \Biggl\{
        -\frac{256}{9} L^3 \biggl[
                \frac{5}{2} (1-x)
                +(x+1) H_0
        \biggr]
        +\frac{32}{3} L^2 \biggl[
                (1-x) \biggl(
                         5 H_1
                        -\frac{7}{3}
                \biggl)
\nonumber \\ &&
                +2 (x+1) \biggl(
                        \frac{1}{2} H_0^2
                        +H_{0,1}
                        -\zeta_2
                \biggr)
                -\frac{1}{3} (19 x+1) H_0
        \biggr]
        +\frac{32}{3} L \biggl\{
                (1-x) \biggl[
                        -\frac{328}{9}
\nonumber \\ &&
                        +\biggl(
                                \frac{2}{3}
                                +10 H_0
                        \biggr) H_1
                        -5 H_1^2
                        -10 H_{0,1}
                \biggr]
                +(x+1) \biggl[
                        4 \big(
                                H_{0,1}
                                -\zeta_2
                        \big) H_0
                        -\frac{2}{3} H_0^3
\nonumber \\ &&
                        -4 H_{0,0,1}
                        -4 H_{0,1,1}
                        +8 \zeta_3
                \biggr]
                -\frac{x+7}{3} \big(
                        H_0^2
                        -2 H_{0,1}
                        +2 \zeta_2
                \big)
                -\frac{2}{9} (83 x+41) H_0
        \biggr\}
\nonumber \\ &&
        +\frac{32}{3} (1-x) \biggl[
                \frac{556}{27}
                +\biggl(
                        -\frac{116}{9}
                        -2 H_0
                        +5 H_0^2
                        -5 \zeta_2
                \biggr) H_1
                +\frac{2}{3} H_1^2
                -10 H_0 H_{0,1}
\nonumber \\ &&
                -\frac{5}{6} H_1^3
                +2 H_{0,1}
                +10 H_{0,0,1}
        \biggr]
        +\frac{32}{9} H_0 \biggl[
                \frac{2}{9} (353 x+17)
                -6 (3 x-1) H_{0,1}
\nonumber \\ &&
                +(19 x+1) \zeta_2
        \biggr]
        +\frac{32}{3} (x+1) \biggl[
                4 \biggl(
                        -H_{0,0,1}
                        +\frac{8}{3} \zeta_3
                \biggr) H_0
                +\big(
                        2 H_{0,1}
                        -\zeta_2
                \big) H_0^2
\nonumber \\ &&
                -\frac{1}{12} H_0^4
                -2 H_{0,1,1,1}
                -2 H_{0,1} \zeta_2
                +\frac{14}{5} \zeta_2^2
        \biggr]
        -16 (5 x+1) H_0^2
        +\frac{16}{9} (3 x-1) H_0^3
\nonumber \\ &&
        -\frac{128}{27} (14 x+11) H_{0,1}
        +\frac{128}{3} (3 x-1) H_{0,0,1}
        +\frac{64}{9} (5 x+2) H_{0,1,1}
\nonumber \\ &&
        +\frac{160}{27} (7 x+13) \zeta_2
        -\frac{64}{9} (33 x-14) \zeta_3
\Biggl\}
+\textcolor{blue}{C_F T_F^2 N_F} \Biggl\{
        -\frac{32}{9} L^3 \big[
                5 (1-x)
\nonumber \\ &&
                +2 (x+1) H_0
        \big]
        +\frac{32}{3} L^2 \biggl[
                (1-x) \biggl(
                        \frac{1}{3}
                        -5 H_1
                \biggr)
                +(x+1) \big(
                        H_0^2
                        -2 H_{0,1}
                        +2 \zeta_2
                \big)
\nonumber \\ &&
                +\frac{x+7}{3} H_0
        \biggr]
        +\frac{32}{3} L \biggl\{
                (1-x) \biggl[
                         \biggl(
                                 10 H_0
                                -\frac{2}{3}
                        \biggr) H_1
                        -\frac{212}{9}
                        -\frac{5}{2} H_1^2
                        -10 H_{0,1}
                \biggr]
\nonumber \\ &&
                +(x+1) \biggl[
                        \big(
                                4 H_{0,1}
                                -4 \zeta_2
                        \big) H_0
                        -\frac{2}{3} H_0^3
                        -4 H_{0,0,1}
                        -2 H_{0,1,1}
                        +6 \zeta_3
                \biggr]
\nonumber \\ &&
                -\frac{2}{9} (55 x+19) H_0
                -\frac{x+7}{3} H_0^2
                +\frac{2}{3} (4 x-5) \big(\zeta_2 - H_{0,1}\big)
        \biggr\}
        -16 (5 x+1) H_0^2
\nonumber \\ &&
        +\frac{32}{3} (1-x) \biggl[
                8
                +\biggl(
                        -2 H_0
                        +5 H_0^2
                        +\frac{5 \zeta_2}{2}
                \biggr) H_1
                +2 H_{0,1}
                -10 H_0 H_{0,1}
                +\frac{\zeta_2}{3}
\nonumber \\ &&
                +10 H_{0,0,1}
        \biggr]
        +\frac{64}{3} \biggl[
                -3
                +7 x
                +(1-3 x) H_{0,1}
                +\frac{1}{6} (4 x-5) \zeta_2
        \biggr] H_0
\nonumber \\ &&
        +\frac{32}{3} (x+1) \biggl[
                \biggl(
                        -4 H_{0,0,1}
                        +\frac{26}{3} \zeta_3
                \biggr) H_0
                +\biggl(
                        2 H_{0,1}
                        -\zeta_2
                \biggr) H_0^2
                -\frac{1}{12} H_0^4
\nonumber \\ &&
                +\zeta_2 H_{0,1}
                -\zeta_2^2
        \biggr]
        +\frac{16}{9} (3 x-1) H_0^3
        +\frac{128}{3} (3 x-1) H_{0,0,1}
        -\frac{32}{9} (41 x-17) \zeta_3
\Biggr\}
\nonumber \\ &&
+\textcolor{blue}{C_F^2 T_F} \Biggl\{
        \frac{32}{3} L^3 \biggl[
                -(1-x) \biggl(
                         \frac{13}{8}
                        +\frac{5}{2} H_1
                \biggr)
                +(x+1) \biggl(
                        -\frac{1}{4} H_0^2
                        -H_{0,1}
                        +\zeta_2
                \biggr)
\nonumber \\ &&
                +\frac{1}{2} (3 x-2) H_0
        \biggr]
        +32 L^2 \biggl\{
                \frac{5}{2} (1-x) \biggl(
                         H_{0,1}
                        -\frac{5}{2}
                        -\frac{3}{2} H_1
                        -H_0 H_1
                \biggr)
                -3 H_{0,1}
\nonumber \\ &&
                +(x+1) \biggl[
                        \big(
                                 \zeta_2
                                -H_{0,1}
                        \big) H_0
                        -\frac{1}{6} H_0^3
                        +H_{0,0,1}
                        -\zeta_3
                \biggr]
                +\frac{1}{4} (7 x-19) H_0
                +3 \zeta_2
\nonumber \\ &&
                +\frac{1}{2} (2 x-3) H_0^2
        \biggr\}
        +32 L \biggl\{
                (1-x) \biggl[
                        \frac{27}{4}
                        +\biggl(
                                \frac{47}{4} H_1
                                +\frac{5}{2} H_1^2
                        \biggr) H_0
                        +\frac{13}{12} H_0^3
\nonumber \\ &&
                        +\biggl(
                                \frac{59}{4}
                                -5 H_{0,1}
                        \biggr) H_1
                        -\frac{5}{8} H_1^2
                        +\frac{5}{12} H_1^3
                        -\frac{47}{4} H_{0,1}
                        +5 H_{0,1,1}
                \biggr]
                -17 H_{0,0,1}
\nonumber \\ &&
                +(x+1) \biggl[
                        \big(
                                3 H_{0,0,1} 
                                +2 H_{0,1,1}
                                +6 \zeta_3
                        \big) H_0
                        +\frac{3}{16} H_0^4
                        -H_{0,1}^2
                        -\frac{3}{2} H_{0,1,1}
                        -3 \zeta_2 H_0^2
\nonumber \\ &&
                        -3 H_{0,0,0,1}
                        +H_{0,1,1,1}
                        +\frac{9}{5} \zeta_2^2
                \biggr]
                +\frac{7}{2} (2 x+1) \big(
                        H_{0,1}
                        -\zeta_2
                \big)
                +\biggl[
                        \frac{1}{8} (87-115 x)
\nonumber \\ &&
                        -3 (x-3) H_{0,1}
                        +(6 x-1) \zeta_2
                \biggr] H_0
                +\frac{1}{16} (87-17 x) H_0^2
                +\frac{1}{2} (67-27 x) \zeta_3
        \biggr\}
\nonumber \\ &&
        +32 (1-x) \biggl\{
                 \biggl(
                         \frac{5}{2} H_{0,0,1}
                        -\frac{35}{4} H_{0,1}
                        +10 H_{0,1,1}
                \biggr) H_0
                -\frac{5}{4} H_0^2 H_{0,1}
                -\frac{5}{2} H_{0,1}^2
\nonumber \\ &&
                +\biggl[
                         \frac{35}{8} H_0^2
                        +\frac{5}{12} H_0^3
                        +\frac{9}{4}
                        -H_{0,1}
                        +\biggl(
                                \frac{81}{4}
                                -5 H_{0,1}
                                +\frac{5}{2} \zeta_2
                        \biggr) H_0
                        +10 H_{0,0,1}
\nonumber \\ &&
                        +10 H_{0,1,1}
                        +\frac{37}{8} \zeta_2
                        -\frac{55}{6} \zeta_3
                \biggr] H_1
                -\frac{83}{2}
                +\biggl(
                        \frac{35}{4}
                        -\frac{5}{2} H_{0,1} 
                        +\frac{25}{8} \zeta_2
                \biggr) H_1^2
                +\frac{3}{4} H_1^3
\nonumber \\ &&
                +\frac{5}{48} H_1^4
                -\biggl(
                         \frac{81}{4}
                        +\frac{5}{2} \zeta_2
                \biggr) H_{0,1}
                +\frac{35}{4} H_{0,0,1}
                +2 H_{0,1,1}
                -\frac{5}{2} H_{0,0,0,1}
                -10 H_{0,0,1,1}
\nonumber \\ &&
                -15 H_{0,1,1,1}
        \biggr\}
        +8 \biggl[
                -29
                -134 x
                +(43 x+25) H_{0,1}
                -\frac{1}{4} (255 x+83) \zeta_2
\nonumber \\ &&
                -28 (x+1) H_{0,0,1}
                +\frac{2}{3} (3 x-16) \zeta_3
        \biggr] H_0
        +8 \big[
                16
                -79 x
                -(23 x-15) \zeta_2
        \big] H_{0,1}
\nonumber \\ &&
        +2 \big[
                1
                +27 x
                +8 (4 x+5) H_{0,1}
                -2 (3 x-7) \zeta_2
        \big] H_0^2
        -\frac{2}{3} (83 x+1) H_0^3
\nonumber \\ &&
        +32 (x+1) \biggl[
                \biggl(
                         5 H_{0,0,0,1}
                        -H_{0,1}^2
                        +4 H_{0,0,1,1}
                        +\zeta_2 H_{0,1}
                        -\frac{14}{5} \zeta_2^2
                \biggr) H_0
                +\frac{1}{240} H_0^5
\nonumber \\ &&
                -\biggl(
                         \frac{3}{2} H_{0,0,1}
                        +\frac{11}{12} \zeta_3
                \biggr) H_0^2
                +\biggl(
                        \frac{1}{6} H_{0,1}
                        +\frac{\zeta_2}{8}
                \biggr) H_0^3
                -6 H_{0,0,0,0,1}
                -24 H_{0,0,0,1,1}
\nonumber \\ &&
                +\biggl(
                        4 H_{0,0,1}
                        -2 H_{0,1,1}
                        -\frac{11}{3} \zeta_3
                \biggr) H_{0,1}
                -10 H_{0,0,1,0,1}
                +13 H_{0,0,1,1,1}
                +\zeta_2 H_{0,0,1}
\nonumber \\ &&
                +6 H_{0,1,0,1,1}
                +H_{0,1,1,1,1}
                +\frac{5}{2} \zeta_2 H_{0,1,1}
                -\frac{5}{6} \zeta_2 \zeta_3
                +5 \zeta_5
        \biggr]
        -\frac{2}{3} (x-3) H_0^4
\nonumber \\ &&
        +16 (5 x-1) H_{0,1}^2
        -16 (28 x+19) H_{0,0,1}
        +8 (3 x+50) H_{0,1,1}
\nonumber \\ &&
        +96 (3 x+2) H_{0,0,0,1}
        -32 (9 x-11) H_{0,0,1,1}
        -16 (3 x-7) H_{0,1,1,1}
\nonumber \\ &&
        +(926 x-422) \zeta_2
        +\frac{8}{5} (23 x-163) \zeta_2^2
        +\frac{4}{3} (305 x-59) \zeta_3
\Biggr\}
\nonumber \\ &&
+\textcolor{blue}{C_A C_F T_F} \Biggl\{
        \frac{32}{3} L^3 \biggl[
                (1-x) \biggl(
                        \frac{223}{12}
                        +\frac{5}{2} H_1
                \biggr)
                +(x+1) \big(
                        H_{0,1}
                        -\zeta_2
                \big)
\nonumber \\ &&
                +\frac{1}{6} (38 x+53) H_0
                +\frac{1}{2} (2-x) H_0^2
        \biggr]
        +32 L^2 \biggl\{
                (1-x) \biggl[
                        -\frac{1}{3} H_0^3
                        +\frac{55}{12} H_1
\nonumber \\ &&
                        +\frac{739}{36}
                        -H_0 \biggl(
                                \frac{5}{2} H_1
                                +2 H_{0,-1}
                                +\zeta_2
                        \biggr)
                        +\frac{5}{2} H_{0,1}
                        +4 H_{0,0,-1}
                \biggr]
                +2 (x-2) \zeta_3
\nonumber \\ &&
                +(x+1) \biggl[
                       -\big(
                                 5 H_{-1}
                                +H_{0,1}
                        \big) H_0
                        +5 H_{0,-1}
                        +H_{0,0,1}
                \biggr]
                +\frac{1}{36} (703 x-11) H_0
\nonumber \\ &&
                -\frac{1}{12} (50 x+29) H_0^2
                +\frac{1}{6} (17 x+5) H_{0,1}
                -\frac{1}{6} (17 x+35) \zeta_2
        \biggr\}
\nonumber \\ &&
        +32 L \biggl\{
                (1-x) \biggl[
                         \biggl(
                                 4 H_{0,-1,-1}
                                -4 H_{0,0,-1}
                                -\frac{265}{6} H_1
                                -\frac{15}{4} H_1^2
                                -\frac{33}{4} H_{0,1}
                        \biggr) H_0
\nonumber \\ &&
                        +\frac{5023}{27}
                        +\biggl(
                                \frac{33}{8} H_1
                                +H_{0,-1}
                                +\frac{\zeta_2}{4}
                        \biggr) H_0^2
                        +\biggl(
                                -\frac{451}{18}
                                +5 H_{0,1}
                                -\frac{5}{2} \zeta_2
                        \biggr) H_1
\nonumber \\ &&
                        +\frac{1}{24} H_1^2
                        -\frac{5}{12} H_1^3
                        -2 H_{0,-1}^2
                        +\frac{265}{6} H_{0,1}
                        +\frac{33}{4} H_{0,0,1}
                        -\frac{5}{2} H_{0,1,1}
                        +6 H_{0,0,0,-1}
\nonumber \\ &&
                        +2 \zeta_2 H_{0,-1}
                \biggr]
                +(x+1) \biggl[
                        \biggl(
                                -
                                \frac{51}{2} H_0
                                +3 H_0^2
                                -10 H_{0,-1}
                                +2 H_{0,1}
                                +3 \zeta_2
                        \biggr) H_{-1}
\nonumber \\ &&
                        -\big[
                                 6 H_{0,-1}
                                +7 H_{0,0,1}
                                +3 H_{0,1,1}
                        \big] H_0
                        +5 H_{-1}^2 H_0
                        +\frac{51}{2} H_{0,-1}
                        +\frac{7}{4} H_0^2 H_{0,1}
\nonumber \\ &&
                        +H_{0,1}^2
                        +10 H_{0,-1,-1}
                        -2 H_{0,-1,1}
                        +6 H_{0,0,-1}
                        -2 H_{0,1,-1}
                        +11 H_{0,0,0,1}
\nonumber \\ &&
                        +H_{0,0,1,1}
                        -H_{0,1,1,1}
                        -\zeta_2 H_{0,1}
                \biggr]
                +\biggl[
                         \frac{1}{2} (19 x-13) H_{0,-1}
                        -\frac{1}{6} (82 x+127) H_{0,1}
\nonumber \\ &&
                        +(1-5 x) H_{0,0,-1}
                        +2 (x-2) H_{0,0,1}
                        +\frac{1}{12} (29 x+5) \zeta_2
                        -(9 x+16) \zeta_3
\nonumber \\ &&
                        +\frac{1}{108} (16603 x+5227)
                \biggr] H_0
                +\frac{1}{72} (625-1316 x) H_0^2
                +\frac{1}{36} (100 x+61) H_0^3
\nonumber \\ &&
                -\frac{1}{36} (179 x+557) H_{0,1}
                +\frac{1}{36} (179 x-361) \zeta_2
                +\frac{1}{12} (299 x+425) H_{0,0,1}
\nonumber \\ &&
                +(13-19 x) H_{0,0,-1}
                +3 (5 x-1) H_{0,0,0,-1}
                -\frac{1}{20} (87 x+209) \zeta_2^2
\nonumber \\ &&
                +\frac{1}{3} (7 x-2) H_{0,1,1}
                -6 (x-2) H_{0,0,0,1}
                +\frac{1}{24} (4-5 x) H_0^4
                +\frac{1}{2} (4 x-137) \zeta_3
        \biggr\}
\nonumber \\ &&
        +8 \biggl(
                8
                -13 x
                -\frac{1}{3} (20 x+17) \zeta_2
        \biggr) H_{0,1} 
        +32 (1-x) \biggl\{
                 \biggl[
                         \frac{7}{2} H_1^2
                        +\frac{5}{6} H_1^3
                        +H_{0,-1}^2
\nonumber \\ &&
                        +\frac{107}{3} H_{0,1}
                        -\frac{25}{2} H_{0,1,1}
                        +\biggl(
                                \frac{263}{6}
                                +\frac{15}{2} H_{0,1}
                                +\frac{5 \zeta_2}{4}
                         \biggr) H_1
                        +4 H_{0,-1,-1,-1}
\nonumber \\ &&
                        -2 H_{0,-1,0,1}
                        -3 H_{0,0,0,-1}
                        -2 H_{0,0,0,1}
                        +\frac{3}{2} \zeta_2 H_{0,-1}
                \biggr] H_0
                +\frac{1}{120} H_0^5
                -\frac{1331}{6}
\nonumber \\ &&
                +\biggl(
                        -\frac{107}{6} H_1
                        -\frac{5}{8} H_1^2
                        -H_{0,-1,-1}
                        +H_{0,0,-1}
                \biggr) H_0^2
                +\biggl(
                        -\frac{1}{6} H_{0,-1}
                        +\frac{\zeta_2}{4}
                \biggr) H_0^3
\nonumber \\ &&
                +\biggl(
                        -\frac{27}{4}
                        -\frac{25}{2} H_{0,0,1}
                        -\frac{5}{2} H_{0,1,1}
                        -\frac{115}{24} \zeta_2
                        +\frac{25}{6} \zeta_3
                \biggr) H_1
                -\biggl(
                         \frac{5}{2}
                        +\frac{5}{8} \zeta_2
                \biggr) H_1^2
\nonumber \\ &&
                -\frac{11}{12} H_1^3
                -\frac{5}{48} H_1^4
                +\big(
                         4 H_{0,0,1}
                        -4 H_{0,-1,-1}
                        -2 H_{0,0,-1}
                        -3 \zeta_3
                \big) H_{0,-1}
                +\frac{5}{2} H_{0,1}^2
\nonumber \\ &&
                -\biggl(
                         \frac{263}{6}
                        +\frac{5}{4} \zeta_2
                \biggr) H_{0,1}
                -\frac{107}{3} H_{0,0,1}
                -7 H_{0,1,1}
                +\frac{25}{2} H_{0,0,1,1}
                +8 H_{0,-1,0,-1,-1}
\nonumber \\ &&
                +\frac{5}{2} H_{0,1,1,1}
                +16 H_{0,0,-1,-1,-1}
                +2 H_{0,0,-1,0,-1}
                -4 H_{0,0,-1,0,1}
                +6 H_{0,0,0,-1,-1}
\nonumber \\ &&
                -12 H_{0,0,0,-1,1}
                +4 H_{0,0,0,0,-1}
                +8 H_{0,0,0,0,1}
                -12 H_{0,0,0,1,-1}
                -4 H_{0,0,1,0,-1}
\nonumber \\ &&
                +2 H_{0,-1,-1} \zeta_2
                -3 H_{0,0,-1} \zeta_2
        \biggl\}
        +32 (x+1) \biggl\{
                \biggl[
                         \frac{7}{2} H_0^2
                        -\frac{5}{12} H_0^3
                        +10 H_{0,-1,-1}
\nonumber \\ &&
                        -5 H_{0,0,-1}
                        +10 H_{0,0,1}
                        +14 H_{0,-1}     
                        +\biggl(
                                21
                                +5 H_{0,-1}
                                -5 H_{0,1}
                                +\frac{15}{4} \zeta_2
                         \biggr) H_0
\nonumber \\ &&
                        -7 \zeta_2
                        -\frac{15}{2} \zeta_3
                \biggr] H_{-1}
                +\biggl(
                        -7 H_0
                        -\frac{5}{4} H_0^2
                        -5 H_{0,-1}
                        +\frac{5}{2} \zeta_2
                \biggr) H_{-1}^2
                +\frac{5}{3} H_{-1}^3 H_0
\nonumber \\ &&
                +\biggl(
                        -7 H_{0,-1}
                        +5 H_{0,-1,1}
                        +\frac{3}{2} H_{0,1}^2
                        -5 H_{0,-1,-1}
                        -\frac{5}{2} H_{0,0,-1}
                        +5 H_{0,1,-1}
\nonumber \\ &&
                        -4 H_{0,0,1,1}
                        +2 H_{0,1,1,1}
                        +\frac{1}{2} \zeta_2 H_{0,1}
                \biggr) H_0
                +\biggl(
                        -\frac{1}{2} H_{0,1,1}
                        +\frac{5}{4} H_{0,-1}
                \biggr) H_0^2
\nonumber \\ &&
                -\biggl(
                         21
                        +\frac{15}{4} \zeta_2
                \biggr) H_{0,-1}
                +\biggl(
                         \frac{5}{3} \zeta_3
                        -5 H_{0,0,1}
                \biggr) H_{0,1}
                -14 H_{0,-1,-1}
                +7 H_{0,0,-1}
\nonumber \\ &&
                -10 H_{0,-1,-1,-1}
                -5 H_{0,-1,0,1}
                +5 H_{0,0,-1,-1}
                -10 H_{0,0,-1,1}
                +27 H_{0,0,0,1,1}
\nonumber \\ &&
                +\frac{5}{2} H_{0,0,0,-1}
                -10 H_{0,0,1,-1}
                +9 H_{0,0,1,0,1}
                -4 H_{0,0,1,1,1}
                -H_{0,1,0,1,1}
                -\frac{\zeta_2}{2} H_{0,0,1}
\nonumber \\ &&
                -\frac{\zeta_2}{2} H_{0,1,1}
                -H_{0,1,1,1,1}
        \biggr\}
        +\frac{32}{3} \biggl[
                 (75 x+41) H_{0,1}
                -12 H_{0,-1,-1}
                -12 H_{0,0,-1}
\nonumber \\ &&
                -3 (x-6) H_{0,-1}
                +2 (22 x+49) H_{0,0,1}
                +6 H_{0,1,1}
                -18 (x-2) H_{0,0,0,1}
\nonumber \\ &&
                +\frac{1}{24} (269-1009 x) \zeta_2
                -\frac{1}{6} (431 x+221) \zeta_3
                -\frac{1}{4} (1993 x+600)
\nonumber \\ &&
                -\frac{3}{10} (5 x-38) \zeta_2^2
        \biggr] H_0
        +32 \biggl[
                \frac{1}{8} (217 x-16)
                +H_{0,-1}
                -\frac{1}{12} (79 x+109) H_{0,1}
\nonumber \\ &&
                +(x-2) H_{0,0,1}
                +\frac{1}{12} (41 x+20) \zeta_2
                +\frac{1}{6} (13 x-14) \zeta_3
        \biggr] H_0^2
        +64 H_{0,-1}^2
\nonumber \\ &&
        +8 (3 x-2) H_{0,1,1}
        +64 (x-6) H_{0,0,-1}
        -\frac{4}{9} (39 x-38) H_0^3
        +192 H_{0,0,0,-1}
\nonumber \\ &&
        +\frac{2}{9} (26 x+23) H_0^4
        -\frac{64}{3} (78 x+41) H_{0,0,1}
        -48 (3 x+29) H_{0,0,0,1}
\nonumber \\ &&
        -16 (3 x+8) H_{0,0,1,1}
        -16 (x+5) H_{0,1,1,1}
        +384 (x-2) H_{0,0,0,0,1}
\nonumber \\ &&
        +\frac{4}{9} (1735 x-133) \zeta_2
        +\frac{8}{3} (37 x+43) \zeta_2 \zeta_3
        +\frac{8}{15} (229 x+1402) \zeta_2^2
\nonumber \\ &&
        -64 \zeta_2 H_{0,-1}
        +\frac{8}{9} (2014 x+851) \zeta_3
        -80 (3 x-7) \zeta_5
\Biggr\}
+a_{Qq}^{(3), \rm PS}(x)~.
\end{eqnarray}
\section{Conclusions}     
\label{sec:4}

\vspace*{1mm}
\noindent
We have calculated the polarized three--loop massive OME $A_{Qq}^{\rm PS, (3)}$ in the single mass case. For the treatment 
of
the Dirac matrix $\gamma_5$ we applied the Larin scheme. It is then convenient to use this scheme also in the 
calculation of the associated massless 
Wilson coefficient. After this, an expression for the pure singlet contribution to the structure function 
can be obtained, also referring to parton distribution functions in the Larin scheme, which are obtained in fits describing the 
evolution in the Larin scheme \cite{Moch:2014sna,Behring:2019tus}. More work is needed in the future to construct the transition to 
the $\overline{\sf MS}$ scheme. In this calculation the central quantity is the constant part of the unrenormalized polarized pure 
singlet OME, $a_{Qq}^{\rm PS, (3)}$, since the other contributions to the OME are coming from lower order calculations or are
known from massless calculations to three--loop order. The present massive OME is given by the usual and generalized harmonic sums 
at rational weights in Mellin $N$ space and by harmonic polylogarithms in $x$ space, allowing besides the argument $x$ also 
for the argument $y= 1-2x \in [-1,1]$. The latter functions are obtained from generalized harmonic polylogarithms. Their 
numerical 
representation can be obtained by using the packages of Refs.~\cite{NUM}.
We have calculated the expressions of $a_{Qq}^{\rm PS, (3)}$ in the small and large $x$ regions.
It is in principle possible to calculate the leading contributions of these expressions
in the small and large $x$ region by applying other techniques. To our knowledge that has not been 
done in the 
present case. These terms are not of phenomenological importance, since they receive large corrections form sub-leading 
contributions, which is also known from various other analyses. Therefore, the complete quantity has to be calculated.
The OME $A_{Qq}^{\rm PS, (3)}$ forms one contribution in the polarized variable flavor number scheme at three--loop order.

\vspace*{5mm}   
\noindent
{\bf Acknowledgment.}\\
We thank S.~Klein for collaboration during an early stage of this work. This work was supported in part by the Austrian
Science Fund (FWF) grant SFB F50 (F5009-N15) and has received funding from the European Union's Horizon 2020 research 
and innovation programme under the Marie Sk\/{l}odowska-Curie grant agreement No. 764850, SAGEX, and COST action CA16201: 
Unraveling new physics at the LHC through the precision frontier. The Feynman diagrams have been drawn using {\tt Axodraw} 
\cite{Vermaseren:1994je}.


\begin{thebibliography}{100}
%
\bibitem{Dittmar:2005ed}
  M.~Dittmar {\it et al.},
  {\it Working Group I: Parton distributions: Summary report for the HERA LHC Workshop Proceedings},
  hep-ph/0511119.
%
\bibitem{Boer:2011fh}
  D.~Boer {\it et al.},
  {\it Gluons and the quark sea at high energies: Distributions, polarization, tomography},
  arXiv:1108.1713 [nucl-th].
%
\bibitem{Bethke:2011tr}
  S.~Bethke {\it et al.},
  {\it Workshop on Precision Measurements of $\alpha_s$},
  arXiv:1110.0016 [hep-ph].
%
\bibitem{Moch:2014tta}
  S.~Moch {\it et al.},
  {\it High precision fundamental constants at the TeV scale},
  arXiv:1405.4781 [hep-ph].
%
\bibitem{Alekhin:2016evh}
  S.~Alekhin, J.~Bl\"umlein and S.O.~Moch,
  Mod.\ Phys.\ Lett.\ A {\bf 31} (2016) no.25,  1630023.
%
\bibitem{Buza:1996wv}
  M.~Buza, Y.~Matiounine, J.~Smith and W.~L.~van Neerven,
  Eur.\ Phys.\ J.\ C {\bf 1} (1998) 301--320
  [hep-ph/9612398].
%
\bibitem{Ablinger:2017err}
  J.~Ablinger, J.~Bl\"umlein, A.~De Freitas, A.~Hasselhuhn, C.~Schneider and F.~Wi\ss{}brock,
  Nucl.\ Phys.\ B {\bf 921} (2017) 585--688
  [arXiv:1705.07030 [hep-ph]].
%
\bibitem{Blumlein:2018jfm}
  J.~Bl\"umlein, A.~De Freitas, C.~Schneider and K.~Sch\"onwald,
  Phys.\ Lett.\ B {\bf 782} (2018) 362--366
  [arXiv:1804.03129 [hep-ph]].
%
\bibitem{Bierenbaum:2009mv}
  I.~Bierenbaum, J.~Bl\"umlein and S.~Klein,
  Nucl.\ Phys.\ B {\bf 820} (2009) 417--482
  [arXiv:0904.3563 [hep-ph]];\\
  J.~Bl\"umlein, S.~Klein and B.~T\"odtli,
  Phys.\ Rev.\ D {\bf 80} (2009) 094010
  [arXiv:0909.1547 [hep-ph]].
%
\bibitem{Ablinger:2014nga}
  J.~Ablinger, A.~Behring, J.~Bl\"umlein, A.~De Freitas, A.~von Manteuffel and C.~Schneider,
  Nucl.\ Phys.\ B {\bf 890} (2014) 48--151
  [arXiv:1409.1135 [hep-ph]].
%
\bibitem{Ablinger:2010ty}
  J.~Ablinger, J.~Bl\"umlein, S.~Klein, C.~Schneider and F.~Wi\ss{}brock,
  Nucl.\ Phys.\ B {\bf 844} (2011) 26--54
  [arXiv:1008.3347 [hep-ph]].
%
\bibitem{Ablinger:2014lka}
  J.~Ablinger, J.~Bl\"umlein, A.~De Freitas, A.~Hasselhuhn, A.~von Manteuffel, M.~Round, C.~Schneider and F.~Wi\ss{}brock,
  Nucl.\ Phys.\ B {\bf 882} (2014) 263--288
  [arXiv:1402.0359 [hep-ph]].
%
\bibitem{Behring:2014eya}
  A.~Behring, I.~Bierenbaum, J.~Bl\"umlein, A.~De Freitas, S.~Klein and F.~Wi\ss{}brock,
  Eur.\ Phys.\ J.\ C {\bf 74} (2014) no.9,  3033
  [arXiv:1403.6356 [hep-ph]].
%
\bibitem{Ablinger:2014vwa}
  J.~Ablinger {\it et al.},
  Nucl.\ Phys.\ B {\bf 886} (2014) 733--823
  [arXiv:1406.4654 [hep-ph]].
%
\bibitem{Ablinger:2014uka}
  J.~Ablinger, J.~Bl\"umlein, A.~De Freitas, A.~Hasselhuhn, A.~von Manteuffel, M.~Round and C.~Schneider,
  Nucl.\ Phys.\ B {\bf 885} (2014) 280--317
  [arXiv:1405.4259 [hep-ph]].
%
\bibitem{Ablinger:2017ptf}
  J.~Bl\"umlein, J.~Ablinger, A.~Behring, A.~De Freitas, A.~von Manteuffel and C.~Schneider,
  PoS (QCDEV2017) 031
  [arXiv:1711.07957 [hep-ph]].
%
\bibitem{Ablinger:2018brx}
  J.~Ablinger, J.~Bl\"umlein, A.~De Freitas, A.~Goedicke, C.~Schneider and K.~Sch\"onwald,
  Nucl.\ Phys.\ B {\bf 932} (2018) 129--240
  [arXiv:1804.02226 [hep-ph]].
%
\bibitem{Ablinger:2011pb}
  J.~Ablinger, J.~Bl\"umlein, S.~Klein, C.~Schneider and F.~Wi\ss{}brock,
  arXiv:1106.5937 [hep-ph].
%
\bibitem{Ablinger:2012qj}
  J.~Ablinger, J.~Bl\"umlein, A.~Hasselhuhn, S.~Klein, C.~Schneider and F.~Wi\ss{}brock,
  PoS (RADCOR2011) 031
  [arXiv:1202.2700 [hep-ph]].
%
\bibitem{Ablinger:2019gpu}
  J.~Ablinger, J.~Bl\"umlein, A.~De Freitas, M.~Saragnese, C.~Schneider and K.~Sch\"onwald,
  {\it The three-loop polarized pure singlet operator matrix element with two different masses},
  arXiv:1911.11630 [hep-ph].
%
\bibitem{Behring:2019tus}
  A.~Behring, J.~Bl\"umlein, A.~De Freitas, A.~Goedicke, S.~Klein, A.~von Manteuffel, C.~Schneider and K.~Sch\"onwald,
  Nucl.\ Phys.\ B {\bf 948} (2019) 114753
  [arXiv:1908.03779 [hep-ph]].
%
\bibitem{Buza:1995ie}
  M.~Buza, Y.~Matiounine, J.~Smith, R.~Migneron and W.~L.~van Neerven,
  Nucl.\ Phys.\ B {\bf 472} (1996) 611--658
  [hep-ph/9601302].
%
\bibitem{LCE}
  K.G.~Wilson,
  Phys.\ Rev.\  {\bf 179} (1969) 1499--1512;\\
  R.A.~Brandt,
  Fortsch.\ Phys.\  {\bf 18} (1970) 249--283;\\
  W.~Zimmermann, {\sf Lect. on Elementary Particle Physics and Quantum Field Theory}, Brandeis Summer Inst., Vol.~1, (MIT Press, Cambridge, 1970),~p. 395;\\
  Y.~Frishman,
  Annals Phys.\  {\bf 66} (1971) 373--389.
%
\bibitem{Gross:1973ju}
  D.J.~Gross and F.~Wilczek,
  Phys.\ Rev.\ D {\bf 8} (1973) 3633--3652.
%
\bibitem{Georgi:1951sr}
  H.~Georgi and H.D.~Politzer,
  {Phys.\ Rev.}\ D {\bf 9} (1974) 416--420.
%
\bibitem{Sasaki:1975hk}
  K.~Sasaki,
  {Prog.\ Theor.\ Phys.}\  {\bf 54} (1975) 1816--1827.
%
\bibitem{Ahmed:1975tj}
  M.A.~Ahmed and G.G.~Ross,
  {Phys.\ Lett.}\ B {\bf 56} (1975) 385--390.
%
\bibitem{Altarelli:1977zs}
  G.~Altarelli and G.~Parisi,
  {Nucl.\ Phys.}\ B {\bf 126} (1977) 298--318.
%
\bibitem{Floratos:1977au}
  E.G.~Floratos, D.A.~Ross and C.T.~Sachrajda,
  Nucl.\ Phys.\ B {\bf 129} (1977) 66--88
  Erratum: [Nucl.\ Phys.\ B {\bf 139} (1978) 545--546].
%
\bibitem{Curci:1980uw}
  G.~Curci, W.~Furmanski and R.~Petronzio,
  Nucl.\ Phys.\ B {\bf 175} (1980) 27--92.
%
\bibitem{GonzalezArroyo:1979df}
  A.~Gonzalez-Arroyo, C.~Lopez and F.~J.~Yndurain,
  Nucl.\ Phys.\ B {\bf 153} (1979) 161--186.
%
\bibitem{Moch:1999eb}
  S.~Moch and J.A.M.~Vermaseren,
  {Nucl.\ Phys.}\ B {\bf 573} (2000) 853--907
  [hep-ph/9912355].   
%
\bibitem{Mertig:1995ny}
  R.~Mertig and W.L.~van Neerven,
  Z.\ Phys.\ C {\bf 70} (1996) 637--654
  [hep-ph/9506451v2].
%
\bibitem{SP_PS1}
  W.~Vogelsang,
  Phys.\ Rev.\  D {\bf 54} (1996) 2023--2029
  [hep-ph/9512218];
  Nucl.\ Phys.\  B {\bf 475} (1996) 47--72
  [hep-ph/9603366].
%
\bibitem{Moch:2014sna}
  S.~Moch, J.A.M.~Vermaseren and A.~Vogt,
  Nucl.\ Phys.\ B {\bf 889} (2014) 351--400
  [arXiv:1409.5131 [hep-ph]].
%
\bibitem{POL19}
I. Bierenbaum J. Bl\"umlein, A. De Freitas, S. Klein and K. Sch\"onwald, DESY 15--004, DO--TH 15/01. 
%
\bibitem{Hasselhuhn:2013swa}
  A.~Hasselhuhn,
  {\sf 3-Loop Contributions to Heavy Flavor Wilson Coefficients of Neutral and Charged Current DIS}, 
  DESY-THESIS-2013-050.
%
\bibitem{Buza:1996xr}
  M.~Buza, Y.~Matiounine, J.~Smith and W.~L.~van Neerven,
  Nucl.\ Phys.\ B {\bf 485} (1997) 420--456
  [hep-ph/9608342].
%
\bibitem{Nogueira:1991ex}
  P.~Nogueira,
  J.\ Comput.\ Phys.\  {\bf 105} (1993) 279--289.
%
\bibitem{Larin:1993tq}
  S.A.~Larin,
  Phys.\ Lett.\ B {\bf 303} (1993) 113--118
  [hep-ph/9302240].
%
\bibitem{HVBM}
  G.~'t Hooft and M.J.G.~Veltman,
  Nucl.\ Phys.\  B {\bf 44} (1972) 189--213;\\
  D.A.~Akyeampong and R.~Delbourgo,
  Nuovo Cim.\  A {\bf 17} (1973) 578--586;
A {\bf 18} (1973) 94--104;
A {\bf 19} (1974) 219--224;\\
  P.~Breitenlohner and D.~Maison,
  Commun.\ Math.\ Phys.\  {\bf 52} (1977) 39--54; 55--75.
%
\bibitem{FINREN}
  D.~Kreimer,
  Phys.\ Lett.\ B {\bf 237} (1990) 59--62;\\
  J.G.~K\"orner, D.~Kreimer and K.~Schilcher,
  Z.\ Phys.\ C {\bf 54} (1992) 503--512;\\
D.~Kreimer, {\sf Dimensionale Regularisierung im Standardmodell}, PhD Thesis, U. Mainz (1992).\\
  D.~Kreimer,
  {\it The Role of $\gamma_5$ in dimensional regularization},
  hep-ph/9401354;\\
  E.~Kraus,
  Annals Phys.\  {\bf 262} (1998) 155--259
  [hep-th/9709154];\\
  S.~Weinzierl,
  {\it Equivariant dimensional regularization},
  hep-ph/9903380;\\
  D.~St\"ockinger,
  {\sf Methoden zur Renormierung supersymmetrischer Eichtheorien in der Wess-Zumino-Eichung und deren Anwendung}, PhD 
  Thesis, TU Karlsruhe (2001);\\
  F.~Jegerlehner,
  Eur.\ Phys.\ J.\ C {\bf 18} (2001) 673--679
  [hep-th/0005255].
%
\bibitem{Matiounine:1998re}
  Y.~Matiounine, J.~Smith and W.L. van Neerven,
  Phys.\ Rev.\ D {\bf 58} (1998) 076002
  [hep-ph/9803439].
%
\bibitem{Tentyukov:2007mu}
  M.~Tentyukov and J.~A.~M.~Vermaseren,
  Comput.\ Phys.\ Commun.\  {\bf 181} (2010) 1419--1427
  [hep-ph/0702279];\\
  J.A.M.~Vermaseren,
  {\it New features of FORM},
  arXiv:math-ph/0010025.
%
\bibitem{REDUZE}
  C.~Studerus,
  Comput.\ Phys.\ Commun.\  {\bf 181} (2010) 1293--1300
  [arXiv:0912.2546 [physics.comp-ph]];\\
  A.~von Manteuffel and C.~Studerus,
  {\it Reduze 2 - Distributed Feynman Integral Reduction},
  arXiv:1201.4330 [hep-ph].
%
\bibitem{FERMAT}
R.H.~Lewis, {\it Computer Algebra System {\tt Fermat}}, {\tt http://home.bway.net/lewis}.
%
\bibitem{Bauer:2000cp}
  C.W.~Bauer, A.~Frink and R.~Kreckel,
  J.\ Symb.\ Comput.\  {\bf 33} (2002) 1--12
  [cs/0004015 [cs-sc]].
%
\bibitem{SIG1}
C.~Schneider, {S\'em.~Lothar. Combin.\/} {\bf 56} (2007) 1--36
 article B56b.
%
\bibitem{SIG2}
C.~Schneider, Simplifying Multiple Sums in Difference Fields, in:~{{\sf Computer
Algebra in Quantum Field Theory: Integration,
  Summation and Special Functions}\/} Texts and Monographs in Symbolic
  Computation eds. C.~Schneider and J.~Bl\"umlein  (Springer, Wien, 2013) 325--360
  [arXiv:1304.4134 [cs.SC]].
%
\bibitem{EMSSP}
  J.~Ablinger, J.~Bl\"umlein, S.~Klein and C.~Schneider,
  Nucl.\ Phys.\ Proc.\ Suppl.\  {\bf 205-206} (2010) 110--115
  [arXiv:1006.4797 [math-ph]];\\
  J.~Bl\"umlein, A.~Hasselhuhn and C.~Schneider,
  PoS (RADCOR 2011) 032
  [arXiv:1202.4303 [math-ph]];\\
  C. Schneider,  
  Computer Algebra Rundbrief {\bf 53} (2013) 8--12;\\
  C.~Schneider,
  J.\ Phys.\ Conf.\ Ser.\  {\bf 523} (2014) 012037
  [arXiv:1310.0160 [cs.SC]].
%
\bibitem{Vermaseren:1998uu}
  J.A.M.~Vermaseren,
  Int.\ J.\ Mod.\ Phys.\ A {\bf 14} (1999) 2037--2076
  [hep-ph/9806280].
%
\bibitem{Blumlein:1998if}
  J.~Bl\"umlein and S.~Kurth,
  Phys.\ Rev.\ D {\bf 60} (1999) 014018
  [hep-ph/9810241].
%
\bibitem{HARMONICSUMS}
  J.~Ablinger,
  PoS (LL2014) 019   [arXiv:1407.6180 [cs.SC]];\\
  {\sf A Computer Algebra Toolbox for Harmonic Sums Related to Particle Physics}, Diploma Thesis, J. Kepler University Linz, 2009,
  arXiv:1011.1176 [math-ph].
%
\bibitem{Ablinger:2011te}
  J.~Ablinger, J.~Bl\"umlein and C.~Schneider,
  J.\ Math.\ Phys.\  {\bf 52} (2011) 102301
  [arXiv:1105.6063 [math-ph]].
%
\bibitem{Ablinger:2013cf}
  J.~Ablinger, J.~Bl\"umlein and C.~Schneider,
  J.\ Math.\ Phys.\  {\bf 54} (2013) 082301
  [arXiv:1302.0378 [math-ph]].
%
\bibitem{Ablinger:2014bra}
  J.~Ablinger, J.~Bl\"umlein, C.G.~Raab and C.~Schneider,
  J.\ Math.\ Phys.\  {\bf 55} (2014) 112301
  [arXiv:1407.1822 [hep-th]].
%
\bibitem{Ablinger:PhDThesis}
  J.~Ablinger,
  {\sf Computer Algebra Algorithms for Special Functions in Particle Physics},
  Ph.D. Thesis, J. Kepler University Linz (2012)
  arXiv:1305.0687 [math-ph].
%
\bibitem{Blumlein:2018cms}
  J.~Bl\"umlein and C.~Schneider,
  Int.\ J.\ Mod.\ Phys.\ A {\bf 33} (2018) no.17,  1830015
  [arXiv:1809.02889 [hep-ph]].
%
\bibitem{Steinhauser:2000ry}
  M.~Steinhauser,
  Comput.\ Phys.\ Commun.\  {\bf 134} (2001) 335--364
  [hep-ph/0009029].
%
\bibitem{Blumlein:2009cf}
  J.~Bl\"umlein, D.J.~Broadhurst and J.A.M.~Vermaseren,
  Comput.\ Phys.\ Commun.\  {\bf 181} (2010) 582--625
  [arXiv:0907.2557 [math-ph]].
%
\bibitem{Remiddi:1999ew}
  E.~Remiddi and J.A.M.~Vermaseren,
  Int.\ J.\ Mod.\ Phys.\ A {\bf 15} (2000) 725--754
  [hep-ph/9905237].
%
\bibitem{Blumlein:1997em}
  J.~Bl\"umlein and A.~Vogt,
  Phys.\ Rev.\ D {\bf 58} (1998) 014020
  [hep-ph/9712546].
%
\bibitem{Blumlein:1996hb}
  J.~Bl\"umlein and A.~Vogt,
  Phys.\ Lett.\ B {\bf 386} (1996) 350--358
  [hep-ph/9606254].
%
\bibitem{Blumlein:1999ev}
  J.~Bl\"umlein,
  Lect.\ Notes Phys.\  {\bf 546} (2000) 42--57
  [hep-ph/9909449].
%
\bibitem{NUM}
  T.~Gehrmann and E.~Remiddi,
  Comput.\ Phys.\ Commun.\  {\bf 141} (2001) 296--312
  [hep-ph/0107173];\\
  J.~Vollinga and S.~Weinzierl,
  Comput.\ Phys.\ Commun.\  {\bf 167} (2005) 177--194
  [hep-ph/0410259];\\
  D.~Maitre,
  Comput.\ Phys.\ Commun.\  {\bf 174} (2006) 222--240
  [hep-ph/0507152];
  Comput.\ Phys.\ Commun.\  {\bf 183} (2012) 846
  [hep-ph/0703052];\\
  J.~Ablinger, J.~Bl\"umlein, M.~Round and C.~Schneider,
  Comput.\ Phys.\ Commun.\  {\bf 240} (2019) 189--201
  [arXiv:1809.07084 [hep-ph]].
%
\bibitem{Vermaseren:1994je}
  J.A.M.~Vermaseren,
  Comput.\ Phys.\ Commun.\  {\bf 83} (1994) 45--58.
\end{thebibliography}
\end{document}